\documentclass[letterpaper,11pt]{article}
\usepackage{jheppub}
\usepackage{graphicx}
\usepackage{subcaption}
\usepackage{xcolor}
\usepackage{framed}
\definecolor{shadecolor}{rgb}{0.90,0.90,0.90} 

\usepackage{amsmath,amssymb,amsfonts,bm,amscd,mathtools}
\usepackage{graphicx}
\usepackage{multirow}
\usepackage{verbatim}
\usepackage{appendix}
\usepackage{fancybox}
\usepackage{array}
\usepackage{url}
\usepackage{float}
\usepackage{tikz}
\usepackage{dsfont}
\usepackage{amsthm}
\usepackage{thmtools} 
\usepackage{thm-restate}

\def\CA{{\mathcal A}}
\def\CB{{\mathcal B}}
\def\CC{{\mathcal C}}
\def\DD{{\mathcal D}}

\def\CI{{\mathcal I}}

\def\CM{{\mathcal M}}
\def\CN{{\mathcal N}}

\def\be{\begin{equation}}
\def\ee{\end{equation}}
\def\bea{\begin{eqnarray}}
\def\eea{\end{eqnarray}}

\newcommand{\ie}{{\tt{IE}}}
\newcommand{\ec}{{\tt{EC}}}

\newtheorem{definition}{Definition}
\newtheorem{lemma}{Lemma}
\newtheorem{remark}{Remark}
\newtheorem{prop}{Proposition}
\newtheorem{conj}{Conjecture}

\title{Analysis of $s$-$t$ symmetric classical S-matrices}

\author{Abhijit Gadde, Shraiyance Jain}
\affiliation{Department of Theoretical Physics \\ 
Tata Institute for Fundamental Research, Mumbai 400005}

\emailAdd{abhijit@theory.tifr.res.in, shraiyance.jain@tifr.res.in}
\preprint{TIFR/TH/25-5}

\abstract{
   We analyze the complex analytic properties of Classical (tree-level) S-matrices for four scalar particles with $s$-$t$ crossing symmetry, involving an infinite number of exchanges. Under suitable analytic conditions, we demonstrate that such S-matrices exhibit a spectrum of poles that is equally spaced. We extend this result to S-matrices with accumulating poles, proving that under analogous conditions, their pole spectrum coincides with that of the Coon S-matrix. The boundedness of the S-matrix in the Regge limit is not essential for our results. While studying S-matrices that do not meet the conditions of our theorems, we encounter functions that have novel non-isolated singularities akin to what is called  the natural boundary.
}

\begin{document}
\maketitle
\flushbottom

\section{Introduction and outlook}
How unique is the classical scattering of gravitons? A sequence of  conjectures \cite{Chowdhury:2019kaq} - called ``classical Regge growth (CRG) conjectures'' -   aims to answer this question.
\begin{enumerate}
    \item {\bf Conjecture 1:} There exist exactly three classical gravitational S-matrices. They are: the S-matrix in pure Einstein gravity, the S-matrix in type II string theory and the S-matrix in heterotic string theory.
    \item {\bf Conjecture 2:} The only consistent classical gravitational S-matrix whose  poles correspond to particles that are bounded in spin is the Einstein S-matrix.
    \item {\bf Conjecture 3:} The only consistent classical gravitational S-matrix with only graviton exchange pole is the Einstein S-matrix. 
\end{enumerate}
As the name suggests, these conjectures are inspired by the bound on the growth of the S-matrix in the so called Regge limit i.e. in the limit of large $s$ with fixed $t$ \cite{Maldacena:2015waa, Camanho:2014apa, Chandorkar:2021viw, Haring:2022cyf}. 
Conjecture $1$ is the strongest and conjecture $3$ is the weakest. The evidence for conjecture $3$, at the four-point level, was provided in \cite{Chowdhury:2019kaq} while conjecture $2$ was examined in \cite{Chakraborty:2020rxf} by classifying three point functions that are quadratic in gravitons. It is hard make progress towards proving conjecture $1$ because, as it turns out, the four point analysis is insufficient to address it. In particular,  it is possible to construct families of seemingly consistent four-point graviton S-matrices from the Virasoro-Shapiro S-matrix by taking linear combinations with shited arguments \cite{Haring:2023The}. Because of this, the classification given in conjecture $1$ is expected to emerge only after  imposing consistency of  higher point amplitudes.

In this paper, we ask a question that is similar in spirit to the exploration of conjecture $1$ but not quite the same. We ask, how unique is the classical S-matrix of open strings? More precisely, how unique is the four point S-matrix of scalars that is $s$-$t$ crossing symmetric and that involves exchange of particles of unbounded spin? 
Ever since Veneziano wrote down the S-matrix  \cite{Veneziano:1968Con} that would become the canonical S-matrix of open strings, the question of its uniqueness has fascinated many \cite{Coon:1969Uni, Baker:1970Dua,  Arkani-Hamed:2022gsa, Cheung:2022Ven}. 
The basic physical principles of analyticity, unitarity and crossing, along with the bound on growth in Regge limit are not enough to constrain the S-matrices of this type. In particular, there is no analogue of conjecture $1$ for open string S-matrices. Seemingly consistent four point amplitudes obeying these conditions were constructed and discussed in \cite{Figueroa:2022Uni, Chakravarty:2022Ont,  Bhardwaj:2023Onu, Cheung:2023Str, Jepsen:2023Cut, Li:2023Tow,  Cheung:2023Bes, Bhardwaj:2023Dua, Haring:2023The,      Rigatos:2024Pos, Eckner:2024The,  Rigatos:2024Coo, Wang:2024Pos,   Bhardwaj:2024klc, Albert:2024yap}. Some of these amplitudes, such as the ``bespoke amplitude'' of \cite{Cheung:2023Bes}, although consistent at the four point level, are shown not to be consistent when generalized to  $n$-points \cite{Arkani-Hamed:2023jwn}.

An alternate line of investigation is to come up with  certain \emph{extra}, reasonably well-motivated conditions on the S-matrix which strongly constrain the $s$-$t$ symmetric S-matrices. 
The problem of discovering conditions - in addition to those of unitarity, crossing and Regge boundedness - that lead to significant constraints on  S-matrices has been considered in the literature. Below we summarize some of the attempts in this direction:
\begin{enumerate}
    \item In \cite{Cheung:2024uhn}, authors imposed a condition on the residue polynomials at successive poles, called ``level-truncation''  and bootstrapped the open string amplitude to be the Veneziano amplitude using this condition. In \cite{Cheung:2024obl}, a similar condition was imposed on closed string S-matrix and a two parameter family of amplitudes was obtained as a solution.
    \item In \cite{Geiser:2022Pro, Geiser:2022Gen}, the zeros of the amplitude were assumed to arise from crossing symmetric polynomials that are linear in $s$  and $t$.  Authors solved this condition to obtain, Veneziano and Coon amplitude for open strings and Virasoro-Shapiro amplitude for the closed string. Some of the other solutions obtained are unphysical as they do not give polynomial residue. 
    \item In \cite{Haring:2023The}, the authors assumed that the spectum is equidistant, obtained a complete parametrization of such S-matrices as a linear sum of Veneziano amplitudes with shifted arguments \cite{PhysRev.185.1876} and constrained the coefficients using unitarity.
    \item In \cite{Huang:2020nqy,Chiang:2023quf}, the authors imposed a certain ``monodromy condition'' that follows from a worldsheet integral representation of the open string amplitude. It relates  S-matrices with different cyclic ordering of external particles. With this condition, authors numerically bootstrap effective field theory to be very close to the one following from Veneziano amplitude.
\end{enumerate}
In the present paper, we follow a similar approach and come up with certain conditions on the analytic structure of the S-matrix which strongly constrain the S-matrix. Following are the two conditions that we consider. 
\begin{enumerate}
    \item {\bf Intrinsic extremality ({\tt IE}):} The zeros of S-matrix in the $s$-plane have bounded derivative with respect to $t$ as $t$ varies  along the real line. 
    \item {\bf Enhanced crossing symmetry ({\tt EC}):} The ``meromorphic part'' of the S-matrix is separately crossing symmetric. 
\end{enumerate}
Our results apply to S-matrices that obey some version of the above conditions. This will be discussed at length in the  bulk of the paper. Let us give a brief motivation for condition $1$ (\ie). 
Physical four point S-matrices form a convex cone i.e.  one can construct a four point S-matrix by taking positive linear combinations of other four point S-matrices. This follows directly from definition \ref{physicalS} of the physical S-matrix. 
We expect such arbitrary linear combinations to not admit a consistent lift to higher points. With this motivation, it is useful to narrow our search down to extremal S-matrices i.e. to S-matrices that can not be written as convex combination of other S-matrices. This program is difficult to carry out in practice because finding extremal S-matrices requires understanding the space of S-matrices to begin with. As we will show in  theorems \ref{ie-finite} and \ref{connection}, the property of intrinsic extremality serves as a good proxy for the property of extremality and moreover, has the advantage of being defined intrinsically. 
Condition $2$ (\ec) requires a longer discussion. We will not have it here. We discuss this condition in detail in section \ref{enhanced}. Let us only comment that we believe that condition $2$ to be true in general and hence,  redundant. Nevertheless, we treat it as an independent condition just so that our results are on firm ground. 
One of our main results is:
\begin{framed}
\begin{itemize}
    \item An S-matrix that is \ie\ and \ec\ has equidistant spectrum.
\end{itemize}
\end{framed}
\noindent
The precise  result is stated as theorem \ref{main-theorem}. We also prove an analogous theorem, theorem \ref{main-theorem2}, for S-matrices whose spectrum has an accumulation point. We show that under similar assumptions,   the poles of such an S-matrix must coincide with those of the Coon amplitude \cite{Coon:1969Uni}.  Curiously the classical Regge growth bound itself does not play any role in deriving above results! So in a sense, our approach is an alternate way of constraining classical S-matrices.  In theorem \ref{regge-theorem}, we prove a structural result about the Regge trajectory after imposing the \ie\ condition.

The reason we consider the $s$-$t$ symmetric S-matrix of scalars rather than $s$-$t$-$u$ symmetric S-matrix of gravitons is its technical simplicity. However the conditions and the approach that we present in this paper, admit a straightforward to graviton S-matrices resulting, possibly, in even  stronger results. We hope to work on this problem in the future. 
Following are some of the other avenues for future research. In this paper, we have exclusively studied four-point amplitudes. It would be interesting to see how the condition of intrinsic extremality and enhanced crossing symmetry be extended to higher points and what constraints on the higher point S-matrix follow.  For this research program, it is important to survey all the open string S-matrices and come up with a unifying principle for all of them.  Interestingly, in \cite{Maldacena:2022ckr} the authors show that open string amplitudes with accumulation points are also possible in string theory. It would be useful to understand the detailed analytic structure of this amplitude and see whether the conditions of intrinsic extremality and enhanced crossing symmetry are obeyed.

Understanding the analytic structure has also been effective in constraining the four-point  correlation function of half-BPS operators in $\CN=4$ $SU(N)$ super Yang-Mills at large $N$ \cite{Alday:2023mvu}. There, the authors express this correlator in Mellin space and obtain conditions on the $AdS$ curvature expansion coefficients using certain expectations, following from string world-sheet integral, about transcendental functions that could appear at a given order. These conditions, along with imposing the correct supergravity limit; the correct structure of poles, and the dimensions of the first few operators, authors construct the first two curvature corrections. It would be interesting to extend the methods presented in this paper to address $AdS$ curvature corrections.

The rest of the paper is organized as follows. In section \ref{general}, we discuss general properties of the four point $s$-$t$ symmetric classical S-matrix. We note that the space of classical S-matrices is a convex cone. We define a weaker notion of the S-matrix called the positive S-matrix. It is the positive S-matrix that we will work with in the rest of the paper. In section \ref{def-res}, we define the additional conditions \ie\ and \ec\ that we impose on the positive S-matrix. We also summarize the main results of the paper in theorems \ref{ie-finite}, \ref{connection}, \ref{main-theorem}, \ref{main-theorem2} and \ref{regge-theorem}. In section \ref{zero-property}, we closely analyze the zeros of the S-matrix. This analysis is crucial in deriving our results. Under certain assumptions we will be able to establish a train-like movement of the zeros along with a degree-increase lemma. Section \ref{intrinsicS} is dedicated to the study of {\tt IE} and partially {\tt IE} S-matrices. We prove theorems \ref{ie-finite}, \ref{connection} and \ref{regge-theorem}. In section \ref{enhanced}, after a brief recollection of Weierstrass factorization theorem, we introduce the canonical form of the S-matrix and discuss in detail the notion of {\tt EC}. We demonstrate that the familiar S-matrices with infinitely many poles viz. the Veneziano amplitude and the Coon amplitude are {\tt EC}. Then we go on to prove theorems \ref{main-theorem} and \ref{main-theorem2}. In section \ref{examples}, we illustrate the validity of our theorems using various examples. Interestingly, we encounter S-matrices whose zero functions have a novel non-isolated singularity.

\section{Classical S-matrix: generalities}\label{general}

In this paper, we aim to characterize four point classical {\emph {a.k.a}} tree-level S-matrices of scalar particles that are symmetric under the crossing of $s$-channel to $t$-channel but not necessarily to $u$-channel. Consequently, it has poles only in $s$ and $t$ channel but not in $u$ channel. Our primary goal is to investigate physical S-matrices arising from the exchange of particles whose spectrum is unbounded in spin. A physical S-matrix is defined by the conditions listed in Definition \ref{physicalS}. Since we exclusively deal with classical S-matrices in this paper, we will omit the adjective ``classical'' in the subsequent discussion.

\begin{framed} 
\begin{definition}[Physical S-matrix in $d$-dimensions]\label{physicalS}
A function ${\cal M}(s,t)$ is called a physical S-matrix in $d$-dimensions if it satisfies the following properties:
\begin{enumerate}
    \item (Reality) ${\CM}(s,t)={\CM}^*(s^*,t^*)$.
    \item (Crossing) $\CM(s,t)= \CM(t,s)$.
\end{enumerate}
The remaining properties are formulated by considering ${\cal M}(s,t)$ as a function $\CM_t(s)$ of $s$ for a fixed value of $t$:
\begin{enumerate}
    \setcounter{enumi}{2}
    \item (Classical) $\CM_t(s)$ is meromorphic, with (possibly infinitely many) poles on the real axis that are independent of $t$ and bounded from below. 
    \item (Unitarity) The residue at any pole $m_i^2$ is a finite sum of $P^{(d)}_\ell(\cos\theta), \cos\theta=1+2t/(m_i^2-4 m_\phi^2)$, with positive coefficients, where $P^{(d)}_\ell(x)$ are $d$-dimensional Gegenbauer polynomials. 
\end{enumerate}
\end{definition}
\end{framed}
\noindent We will label the poles of the S-matrix starting from $m_1^2$ as $m_1^2 <m_2^2 < \ldots$.

\begin{framed}
\begin{remark}
A $d$-dimensional physical S-matrix is also a $d'$-dimensional physical S-matrix for $d'<d$. 
\end{remark}
\end{framed}
\noindent This follows from the fact that $P^{(d)}_\ell(\cos\theta)$ can be expressed as $\sum_{\ell'} c_{\ell'} P^{(d')}_{\ell'}(\cos\theta)$, where $c_{\ell'}>0$ for any $d'<d$. 

Any S-matrix derived from a classical field theory satisfies the defining properties of the physical S-matrix, providing significant flexibility in constructing such S-matrices. This observation is formalized in the following remarks:

\begin{framed}
\begin{remark}\label{convexity}
If $\CM_1(s,t)$ and $\CM_2(s,t)$ are physical S-matrices, then $p_1 \CM_1(s,t) + p_2 \CM_2(s,t)$, with $p_1, p_2 > 0$, is also a physical S-matrix. In other words, physical S-matrices form a convex cone.
\end{remark}
\end{framed}

\begin{framed}
\begin{remark}\label{convexity1}
If $\CM_1(s,t)$ is a physical S-matrix, then so is $\CM_1(s,t) + c \CB(s,t)$, where $c \in \mathbb{R}$, and $\CB(s,t)$ is a physical S-matrix that is an entire function, i.e., one with no poles.
\end{remark}
\end{framed}
\noindent
An example of entire S-matrix $\CB(s,t)$ is an $s$-$t$ symmetric polynomial. A polynomial S-matrix arises due to contact interactions. 
Note that the coefficient $c$ of $\CB(s,t)$ in Remark \ref{convexity1} need not be positive because if an entire function $\CB(s,t)$ is a physical S-matrix, so is $-\CB(s,t)$. This is because, \emph{a priori}, there is no requirement of positivity for the coefficient of the contact terms.\footnote{The dispersive sum rules following from the boundedness of the S-matrix in the Regge limit may put bounds on the contact term coefficients, see \cite{Albert:2024yap} and references therein. However we do not impose any Regge boundedness and disregard the positivity conditions on the contact terms that  follow from it.}

The classification of physical S-matrices is challenging due to remarks \ref{convexity} and \ref{convexity1}. To identify interesting S-matrices with infinitely many poles, such as the open string scattering amplitude or amplitudes in large $N$ gauge theories, an additional property is often imposed on physical S-matrices \cite{Maldacena:2015waa, Camanho:2014apa, Chandorkar:2021viw, Haring:2022cyf}:
\begin{enumerate}
    \setcounter{enumi}{4}
    \item (Classical Regge growth) At large $s$ and for $t \leq 0$, $|\CM_t(s)| \propto |s|^{\alpha}$ with $\alpha \leq 2$.
\end{enumerate} 
In this paper we adopt a slightly different perspective. We will  forego  the Regge growth condition and instead explore other conditions that yield new insights.

\subsection{Motivation from Extremality}
Given that physical S-matrices form a convex cone, it is natural to ask which physical S-matrices are extremal. Extremal physical S-matrices are defined as follows:

\begin{framed}
\begin{definition}[Extremality]
A physical S-matrix $\CM(s,t)$ is extremal if it cannot be expressed as $p_1 \CM_1(s,t) + p_2 \CM_2(s,t)$, with $p_1, p_2 > 0$, where $\CM_i$ are physical S-matrices.
\end{definition}
\end{framed}
\noindent
Unfortunately, Remark \ref{convexity1} implies that no physical S-matrix is extremal. Any physical S-matrix $\CM(s,t)$ can be written as the convex sum:
\begin{align}
    \CM(s,t) = (\CM(s,t) - \CB(s,t)) + \CB(s,t),
\end{align}
where $\CB(s,t)$ is an entire physical S-matrix. 

To salvage this situation, we temporarily tweak the definition of physical S-matrices by requiring that they have at least one pole. Even with this modification, the result is as follows:

\begin{framed}
\begin{prop}\label{extremal1}
An extremal physical S-matrix has exactly one pole. 
\end{prop}
\end{framed}
\begin{proof}
Assume that the extremal S-matrix $\CM(s,t)$ has more than one pole. Let one pole be at $m_1^2$. Then, we can write $\CM(s,t)$ as:
\begin{align}
    \CM(s,t) = \frac{{\rm Res}_{s=m_1^2} P(t)}{s-m_1^2} + \frac{{\rm Res}_{t=m_1^2} P(s)}{t-m_1^2} + \Big(\CM(s,t) - \big(\frac{{\rm Res}_{s=m_1^2} P(t)}{s-m_1^2} + \frac{{\rm Res}_{t=m_1^2} P(s)}{t-m_1^2}\big)\Big).
\end{align}
Each term in this decomposition is a physical S-matrix with at least one pole, contradicting extremality.
\end{proof}

\noindent
While the notion of extremal vectors is essential in finite-dimensional cones, it does not straightforwardly apply to the infinite-dimensional space of S-matrices. For example, the Veneziano amplitude cannot be written as a convergent sum over poles in both the $s$- and $t$-channels. A more nuanced definition of extremality may yet yield intriguing results.

\subsection{Positive S-matrix}
Instead of focusing on extremality, we replace it with the concept of intrinsic extremality, which we find more practical. Before we do so,  we will weaken the definition of physical S-matrices to introduce positive S-matrices. All of our results  use only this weaker definition. 

\begin{framed}
\begin{definition}[Positive S-matrix]\label{positiveS}
A function $\CM(s,t)$ is called a positive S-matrix if it satisfies properties $1$, $2$, and $3$ of the physical S-matrix in Definition \ref{physicalS}, with Property $4$ replaced by:
\begin{enumerate}
    \setcounter{enumi}{3}
    \item (Positivity) The residue at any pole is a polynomial in $t$ that is strictly positive for $t > 4m_\phi^2$.
\end{enumerate}
\end{definition}
\end{framed}
\noindent
We demonstrate why positive S-matrices are weaker than physical S-matrices in proposition \ref{weaker}. Before that, we make several simple observations.
\begin{framed}
\begin{remark}
Positive S-matrices form a convex cone.
\end{remark} 
\end{framed}
\begin{framed}
\begin{remark}\label{scale}
If \(\CM(s,t)\) is a positive S-matrix, then so is \(\CM(4m_\phi^2+\alpha(s-4m_\phi^2), 4m_\phi^2+\alpha(t-4m_\phi^2))\) for \(\alpha > 0\).
\end{remark}
\end{framed}

\noindent
This remark illustrates that a new S-matrix can be obtained by scaling \(s\) and \(t\) with their center at \(4m_\phi^2\). 

\begin{framed}
\begin{remark}\label{translation}
If \(\CM(s,t)\) is a positive S-matrix, then so is \(\CM(s-\beta ,t-\beta)\) for \(\beta > 0\).
\end{remark}
\end{framed}

\noindent
Here, the new S-matrix is derived from the original one by shifting all poles to the right by a constant amount.

As we are not imposing the property of \emph{classical Regge growth} on positive S-matrices, there is a tradeoff. It introduces significant freedom in constructing such S-matrices.

\begin{framed}
\begin{remark}\label{const-zero}
If \(\CM(s,t)\) is a positive S-matrix, then so is \(\CB(s,t)\CM(s,t)\), where \(\CB(s,t)\) is an entire function satisfying:
\begin{itemize}
\item \(\CB(s,t)=\CB^*(s^*,t^*)\)
\item \(\CB(s,t)=\CB(t,s)\)
\item \(\CB(m_i^2, t)\) is a polynomial in \(t\) that is positive for \(t>4m_\phi^2\).
\end{itemize}
\end{remark}
\end{framed}

\noindent
The necessity of the first two properties of \(\CB(s,t)\) is  straightforward. The third property ensures that the residue of the new S-matrix is a polynomial that remains positive for \(t>4m_\phi^2\) as required in the definition of the positive S-matrix. 

A simple example of such a function \(\CB(s,t)\) is \(\prod_{\alpha\in S}(s-\alpha)(t-\alpha)\), where \(S\) contains points that are either real and less than \(4m_\phi^2\) or form complex conjugate pairs. This would produce an S-matrix \(\CM_t(s)\) whose zeros are independent of \(t\). These zeros can always be factored out, and for the rest of the paper, we assume the S-matrix has no constant zeros.

A more involved example of \(\CB(s,t)\) that is not just a polynomial with fixed zeros is \(\exp(\CC(s)\CC(t))\), where \(\CC(s)\) is an entire function with zeros at the poles of \(\CM(s,t)\), i.e., at \(m_i^2\). Thanks to the Weierstrass factorization theorem, such a function can always be constructed, even when the S-matrix has infinitely many poles. The first two properties are clearly satisfied, and the third property holds because \(\CB(m_i^2,t)=1\) by construction.

Although we could simplify matters by imposing the condition of \emph{classical Regge growth}, which eliminates the freedom of multiplying by a non-trivial entire function as outlined in Remark \ref{const-zero}, we choose not to do so. Remarkably, even without imposing this condition, significant constraints on S-matrices, such as those described in Theorems \ref{ie-finite}, \ref{connection}, \ref{main-theorem}, and \ref{main-theorem2}, still emerge. This approach highlights the robustness of the results despite the expanded freedom due to lack of classical Regge growth bound.

Finally, we demonstrate that any physical S-matrix can always be transformed into a positive S-matrix.

\begin{framed}
\begin{prop}\label{weaker}
The function
\begin{align}
\tilde \CM(s,t) \equiv \CM(s,t) \prod_{m_i^2< 4 m_\phi^2} (s-m_i^2)(t-m_i^2),
\end{align}
is a positive S-matrix if \(\CM(s,t)\) is a physical S-matrix. Here, \(m_i^2\) are the poles of the physical S-matrix \(\CM(s,t)\).
\end{prop}
\end{framed}

\noindent
In essence, to obtain a positive S-matrix, we must remove the poles of the physical S-matrix that are smaller than \(4m_\phi^2\).
\begin{proof}
Clearly, \(\tilde \CM(s,t)\) satisfies properties 1, 2, and 3. We only need to verify its positivity. Since all poles of \(\tilde \CM(s,t)\) are greater than \(4m_\phi^2\), we have \(\cos\theta > 1\) for \(t > 0\) at each pole. Gegenbauer polynomials \(P^{(d)}_\ell(\cos\theta)\) are strictly positive for \(\cos\theta > 1\). As the coefficients of \(P^{(d)}_\ell(\cos\theta)\) are positive, the contribution to the residue from \(\CM(s,t)\) is positive for \(t > 0\). However, the contribution from the factor \(\prod_{m_i^2< 4 m_\phi^2} (s-m_i^2)(t-m_i^2)\) is positive definite only for \(t > 4m_\phi^2\). This completes the proof.
\end{proof}

\noindent
From now on, as we work exclusively with positive S-matrices, we will drop the ``positive'' adjective and refer to them simply as S-matrices.

\subsection{Zeros of the S-matrix}\label{zeross}

Most of our analysis of the S-matrix is based on the behavior of the zeros of the S-matrix. By zeros of the S-matrix, we mean the zeros in the complex $s$-plane for a fixed value of $t$. Anaysis of zeros of the classical $s$-$t$ symmetric S-matrix played an important role in \cite{Wanders:1971et, Wanders:1971qf, Caron-Huot:2016icg}. In general the S-matrix has multiple zeros, possibly infinitely many. In fact, when the S-matrix has infinitely many poles, it needs to have infinitely many zeros as we will see. We will refer to these zeros as $f_i(t)$ where the label $i$ is assigned arbitrarily for now. We will make their labeling more specific later on.
In general, the zero function $f_i(t)$ is a function with branch-cuts i.e. it is defined as values of a function on a given sheet of a multi-valued function. The values of the function on the other sheet of the same multi-valued function give rise to another zero on the $s$-plane. At the branch point, multiple zeros coalesce. 

A good model for the zero functions $f_i(t)$ to keep in mind is the zeros of some crossing symmetric polynomial, say $s^2+t^2-1$. Its zeros are $s=\pm \sqrt{1-t^2}$. We can pick $f_1(t)= +\sqrt{1-t^2}$ and  $f_2(t)=- \sqrt{1-t^2}$. Both $f_1$ and $f_2$ have branch points at $t=1, -1$. It is convenient to pick the branch cuts for both the functions that are along the real line. We fix a convention that there is no branch cut at, say $t=0$. This fixes the branch cuts of both $f_1, f_2$ to extend away from the origin after starting at $1$ and $-1$, rather than a single branch cut that connects $1$ and $-1$. For real $t\in (-1,1)$, the two zeros are real while for $t$ outside this region with a small imaginary part $\epsilon$, they are complex conjugates of each other.  
As we cross the branch cut for $f_1$, the function $f_1$ changes discontinuously but that doesn't mean that one of the zeros suddenly jumps in the $s$-plane. But rather, as $t$ approaches the branch cut of $f_1$, it also approaches the branch cut of $f_2$. Both cuts are simultaneously crossed and as a result, the labels on the zeros get switched. The zero labeled by $i=1$ becomes the one labeled by $i=2$ and vice versa. As long as $t$ stays away from the branch cut, the labels on the zero functions are fixed. At the branch-point the two zeros collide on the real axis at $s=0$.

Next consider the polynomial $s^3+t^3-1$ then the zeros are $f_1=\sqrt[3]{1-t^3}, f_2= \omega \sqrt[3]{1-t^3}$ and $\omega^2\sqrt[3]{1-t^3}$ where $\omega=e^{2\pi i/3}$ is the third root of unity. All these functions have three branch points,  at $t=1, \omega$ and $\omega^2$. 
As a matter of convention, for all functions, the branch cuts starting from $\omega$ and $\omega^2$ are oriented so that they do not cross the real axis and are mirror images of each other while the branch cut starting from $t=1$ is oriented along the positive $t$ axis so that they are single valued at $t=0$. As we approach a branch cut of one of the functions, we approach the branch cut of all three functions. After crossing the branch cut, the labels $i=1, 2, 3$ of the zeros get cyclically permuted. As we approach a branch point, all three zeros collide at $s=0$. 

Note that in both these examples, $f_i(t^*)=f_j^*(t)$ for some $j$. For instance, in the first example, $f_i(t^*)=f_i^*(t)$ for $i=1,2$ and in the second example, $f_1(t^*)=f_1^*(t)$ and $f_2(t^*)=f_3^*(t)$. This is not a coincidence. Zeros of any function $\CM(s,t)$ that obeys $\CM(s^*,t^*)=\CM^*(s,t)$ have this property. This is because, if $\CM(s,t)=0$ then of course $\CM(s^*,t^*)=0$. If we write the zero part of\footnote{Such a form can also be written for functions with infinitely many zeros. One has to make sure that this product converges. See section \ref{enhanced} for more discussion about product form of functions with infinitely many zeros.} $\CM(s,t)=\prod_{i=1}(s-f_i(t))$, then it follows that
\begin{framed}
\begin{remark}
$f_i(t^*)=f_j^*(t)$ for some $j$.
\end{remark}
\end{framed}
\noindent 
Let us note another property of the zero functions. The S-matrix $\CM(s,t)$ vanishes at $(s,t)=(f_i(t), t)$ for all $i$ and given that it is crossing symmetric, it vanishes also at $(s,t)=(t,f_i(t))$. Thinking of $f_i(t)$ as a value of $t$, we see that 
\begin{framed}
    \begin{remark}\label{fjisinverse}
    $f_{j}(f_i(t))=t$ for some $j$.
    \end{remark}
    \end{framed}

    \section{Key notions and results}\label{def-res}
    In this section, we will give definitions of the important notions introduced in the paper. We also state the main results about tree-level S-matrix with infinitely many poles that are proved in the body of the paper. 
 
    \subsection{No recombination: outline}
    In order to define no recombination ({\tt NR}) property of the S-matrix, we need to study the movement of zeros on the s-plane as $t$ is varied. This is done in detail in section \ref{zero-property}. In this section, we only outline the definition of property ``no  recombination'' ({\tt NR}) and an analogous but weaker property ``no infinite recombination'' ({\tt NIR}).
    
    As $t$ moves to one of the poles $m_i^2$, all the poles of the S-matrix $\CM_t(s)$ are cancelled by its zeros. This is because, if that were not the case, the residue of the S-matrix at $t=m_i^2$ would have poles in $s$ which contradicts property $4$ of positive S-matrices, see section \ref{zero-property} for details. Moreover, as discussed there, only a single zero cancels each of the the s-poles. It is also shown there that these are the only real zeroes of $\CM_t(s)$ that are greater than $m_1^2$. A real zero of $\CM_t(s)$ remains real as $t$ moves along the real axis unless it ``recombines''  with another real zero and move off of the real axis in a complex conjugate pair. The zeros can leave the real axis only in complex conjugate pairs because of the reality property $\CM(s^*, t^*)=\CM^*(s,t)$. 
    
    Let us denote by $S_i$ the set of the zero functions that cancel all the $s$-poles at $t=m_i^2$. It is also useful to define $\tilde S_i$ to be $S_i$ minus the zero function that is smallest at $t=m_i^2$.  
    As $t$ moves along the real axis\footnote{If there is a branch cut along the real axis for certain zero functions $f_j(t)$ then we assume that the $t$ has a small positive imaginary part.} from $m_i^2$ to $m_{i+1}^2$, some of the zeros can possibly recombine with the zeros of set $\tilde S_i$. This might happen if a complex conjugate pair of the zeros moves to the real axis and one of the zeros of the pair recombines  with a zero of set $\tilde S_i$.
    \begin{framed} 
        \begin{definition}[No recombintation]
            If no zero of $\tilde S_i$ undergoes recombintation as $t$ moves from $m_i^2$ to $m_{i+1}^2$ for all $i$, then we call the S-matrix {\tt NR} (No Recombintation). If only finitely many zeros of $\tilde S_i$ undergo recombintation as $t$ moves from $m_i^2$ to $m_{i+1}^2$ for all $i$, then we call the S-matrix {\tt NIR} (No Infinite Recombintation). 
        \end{definition}
        \end{framed}
        \noindent

\subsection{Intrinsic Extremality}
Let’s return to the discussion of intrinsic extremality. We begin with its definition.

\begin{framed}
\begin{definition}[Intrinsic Extremality]
    An S-matrix is called intrinsically extremal (\ie) in a domain \(\cal D\) if \(|f_i'(t)|\) is bounded within \(\cal D\) for all \(i\). If \(\cal D\) is the entire complex plane (excluding branch points), the S-matrix is called \ie. If \(\cal D\) is an open, connected region containing all the real segments \([m_i^2, m_{i+1}^2]\) for all \(i\), then the S-matrix is referred to as partially intrinsically extremal ({\tt PIE}).
\end{definition}
\end{framed}
\noindent
In simpler terms, the condition of intrinsic extremality ensures that the speed of each zero with respect to changes in \( t \) remains finite. The exclusion of branch points from the \ie\ condition allows S-matrices whose zeros have branch points of the type $f(t)=(t-t_0)^\alpha$ where $\alpha$ is a rational number to be \ie, even though at $t_0$, $f'(t)$ may diverge.
 
This seemingly esoteric notion of intrinsic extremality serves as a proxy for extremality. This is supported by the two theorems that we will prove in section \ref{intrinsicS}.

\begin{shaded*}
\begin{restatable}{thm}{iefinite}\label{ie-finite}
    \ie\ S-matrices with finitely many poles must
    \begin{enumerate}
        \item either be entire functions, 
        \item or have exactly one pole with a constant remainder.
    \end{enumerate}
\end{restatable}
\end{shaded*}
\noindent
Theorem \ref{ie-finite} classifies \ie\ S-matrices with finitely many poles. But what about \ie\ S-matrices with infinitely many poles? 
We show that intrinsic extremality satisfies a property similar in spirit to the defining property of extremality.

\begin{shaded*}
\begin{restatable}{thm}{connection}\label{connection}
    Let \(\CM_\alpha(s,t)\), for \(i=\alpha,\ldots,n\), be \ie\ S-matrices satisfying the following condition:
    \begin{itemize}
        \item Let the set of poles of \(\CM_\alpha\) be \(K_\alpha\), and \(K = \cup_\alpha K_\alpha\). The set \(K \setminus K_\alpha\) is non-empty, and \(\min\{K_\alpha\} < \min\{K \setminus K_\alpha\}\).
    \end{itemize}
    Then the convex combination \(\CM \equiv \sum_{\alpha=1}^n p_\alpha \CM_\alpha\), with \(p_\alpha > 0\), is an S-matrix that is not \ie.
\end{restatable}
\end{shaded*}
\noindent
The condition on the poles may seem daunting. It is simply stating, in a precise manner, that the pole sets of the S-matrices do not overlap completely. In particular, the condition is satisfied for two S-mmatrices whose poles sets are not correlated.

The usefulness of the notion of \ie\ is illustrated by the fact that the Veneziano amplitude is \ie. In this case, the zeros of \(\CM_t(s)\) are simply \(f_i(t) = i - t\) for \(i \in \mathbb{Z}_+\), and $|f_i'(t)|=1<\infty$. In fact, certain linear combinations of Veneziano-like amplitudes considered in \cite{Haring:2023The} are also \ie.  
We do not know whether there are other S-matrices thatx are \ie\, and also if  the S-matrices of large \(N\) gauge theories are \ie. Progress on this front would be valuable.

\subsection{Enhanced Crossing Symmetry: outline}
To make progress in classifying \ie\ S-matrices with infinitely many poles, we introduce a new concept: enhanced crossing symmetry (\ec). It coincides with ordinary crossing symmetry for S-matrices with finitely many poles. It turns out that the Veneziano amplitude is enhanced crossing symmetric (\ec). The definition of enhanced crossing symmetry requires a detailed discussion of the analytic properties of the S-matrix, which we outline below. For more details, please refer to section \ref{enhanced}. We first introduce a canonical form of the S-matrix:

\begin{align}
    \CM(s,t) = \frac{\CN(s,t)}{\DD(s,t)}
\end{align}
where the denominator function is

\begin{align}\label{denom}
    \DD(s,t) = \prod_{n=1}^{\infty}\left(1 - \frac{s}{m_n^2}\right) C_n(s) \prod_{n=1}^{\infty}\left(1 - \frac{t}{m_n^2}\right) C_n(t).
\end{align}
It incorporates all the poles of the S-matrix as zeros. Since there may be infinitely many zeros, the infinite product in equation \eqref{denom} may require a nowhere-vanishing entire function \(C_n(s)\) in each term for convergence, according to the Weierstrass factorization theorem. The denominator \(\DD(s,t)\) is manifestly crossing symmetric. Because \(\CM(s,t)\) is crossing symmetric, the numerator \(\CN(s,t)\) must also be crossing symmetric.

Next, we present the canonical form of the numerator. Consider \(\CN(s,t)\) as an entire function of \(s\) for a fixed \(t\). Using the Weierstrass factorization theorem:

\begin{align}
    N(s,t) = \prod_{n=1}^{\infty} \left(1 - \frac{s}{f_n(t)}\right) B_n(s,t).
\end{align}
Here, \(f_n(t)\) are the zeros of \(\CM_t(s)\), and \(B_n(s,t)\) is a nowhere-vanishing entire function of \(s\). We now introduce the canonical form of the numerator \(N(s,t)\), where we rearrange the product so it can be written in terms of another nowhere-vanishing entire function \(\tilde B_n(s,t)\) of \(s\) that does not have zeros or singularities in \(t\). This canonical form is:

\begin{align}\label{best-form-1}
    N(s,t) = \prod_{n=1}^{\infty}(s - f_n(t)) A_n(t) \tilde B_n(s,t).
\end{align}
Here, \(\tilde B(s,t)\) does not have zeros or singularities in either \(s\) or \(t\). We can fix a convention (see section \ref{enhanced}) for the choice of \(A_n(t)\) such that this canonical form is unique.

\begin{framed}
    \begin{definition}[Enhanced Crossing Symmetry]
        An S-matrix is said to be enhanced crossing symmetric if its canonical form satisfies:
        \begin{align}
           \lim_{N \to \infty} \prod_{n=1}^{N} \frac{\tilde B_n(s,t)}{\tilde B_n(t,s)} = 1.
        \end{align}
    The convergence is uniform in the entire $s$ and $t$ plane. 
    \end{definition}
\end{framed}
\noindent 
Together with the crossing symmetry of the S-matrix, the condition \ec\ implies the following relation:

\begin{align}
    \lim_{N \to \infty} \prod_{n=1}^{N} \frac{A_n(t)(s - f_n(t))}{A_n(s)(t - f_n(s))} = 1.
\end{align}

\subsection{Results}
With the preparatory work complete, we now present our main results:

\begin{shaded*}
\begin{restatable}{thm}{maintheorem}\label{main-theorem}
    An S-matrix that is \ie\ and \ec\ must have a spectrum of poles that is equidistant, i.e.
    \begin{align}
        m_n^2 = b(n-1) + a \qquad \text{for some constants } a, b.
    \end{align}
\end{restatable}
\end{shaded*}

\noindent
In terms of the spectrum of poles, the parameters are \(a = m_1^2\) and \(b = m_2^2 - m_1^2\). We have already established that the Veneziano amplitude is \ie. In section \ref{enhanced}, we show that it is also \ec, and thus its equidistant poles are consistent with theorem \ref{main-theorem}.

Next, we consider the case where the spectrum of poles has an accumulation point, i.e. \(\lim_{n \to \infty} m_n \equiv m_{\infty}\), where \(m_{\infty}\) is finite. Strictly speaking, a function with an accumulation point of poles cannot be an S-matrix because it would not be a meromorphic function of \((s,t)\), but instead would have an essential singularity at the accumulation point. In this case, we impose the milder condition that the S-matrix is meromorphic except at the accumulation point. As theorem \ref{main-theorem} suggests, such an S-matrix cannot be both \ie\ and \ec. However, we can still prove a powerful result for these S-matrices if they are {\tt NIR}.

\begin{shaded*}
\begin{restatable}{thm}{maintheoremtwo}\label{main-theorem2}
    If the poles of an {\tt NIR}, \ec, partially \ie\ S-matrix accumulate, then the poles must satisfy:
    \begin{align}
        m_n^2 = a + b q^{n-1} \qquad \text{for some constants } a, b, q.
    \end{align}
\end{restatable}
\end{shaded*}

\noindent
The parameters \(a\), \(b\), and \(q\) can be written in terms of the spectrum as:
\begin{align}
    a = m_{\infty}^2, \qquad b = m_1^2 - m_{\infty}^2, \qquad q = \frac{m_2^2 - m_{\infty}^2}{m_1^2 - m_{\infty}^2}.
\end{align}
The Coon amplitude satisfies the conditions of theorem \ref{main-theorem2}, and thus its poles follow the expected form.

We now turn to the growth of the S-matrix in the large \(s\), fixed $t$ limit, also known as the Regge limit. The following theorem addresses this growth:

\begin{shaded*}
\begin{restatable}{thm}{reggetheorem}\label{regge-theorem}
    If \(\CM(s,t)\) is \ie\ and if \(\log|\CM_t(s)|/\log|s|\) is finite for \(t > m_1^2\), then at large \(|s|\), we have:
    \begin{align}
        \CM(s,t) = \CA(t) |s|^{j(t)},
    \end{align}
    where \(j(t)\) satisfies the condition: $j(m_n^2)=\ell+n-1$, and
    \begin{align}
        j(m_n^2) < j(t) < j(m_{n+1}^2), \qquad \text{for} \,\, m_n^2 < t < m_{n+1}^2.
    \end{align}
    for some non-negative integer \(\ell\). 
    Additionally, if the \ie\ S-matrix is also {\tt NR}, then \(j(t)\) is completely monotonic for \(t > m_1^2\).
\end{restatable}
\end{shaded*}

\begin{figure}[h]
    \centering
    \includegraphics[scale=0.35]{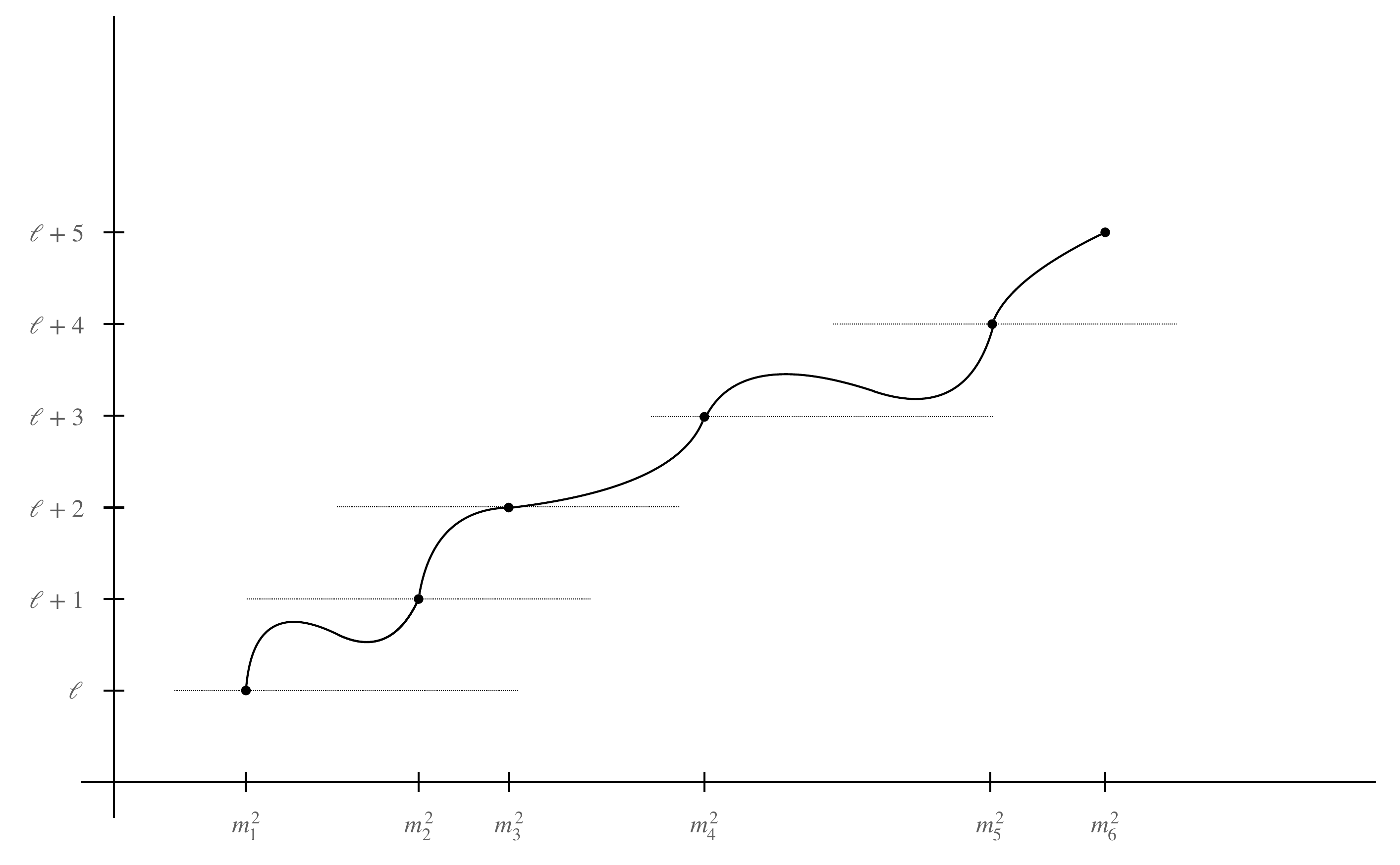}
    \caption{A sample Regge trajectory for an S-matrix obeying the conditions of theorem \ref{regge-theorem}, which is not necessarily NR.}
    \label{regge12}
\end{figure}

\noindent
The function \(j(t)\) is known as the Regge trajectory. A schematic form of the Regge trajectory for an intrinsically extremal S-matrix is shown in figure \ref{regge12}. 

\section{Analysis of Zeros}\label{zero-property}

When necessary, we will refer to the poles and zeros of $\CM_t(s)$ as $s$-poles and $s$-zeros respectively in order to avoid confusion with the poles and zeros of $\CM_s(t)$ which will in turn be referred to as $t$-poles and and $t$-zeros. This will be useful because in the following discussion, we will have to think of $\CM(s,t)$ both as $\CM_s(t)$ and $\CM_t(s)$. Now will collect properties of zeros of $\CM_t(s)$.

\begin{figure}[h]
    \centering
    \includegraphics[scale=0.30]{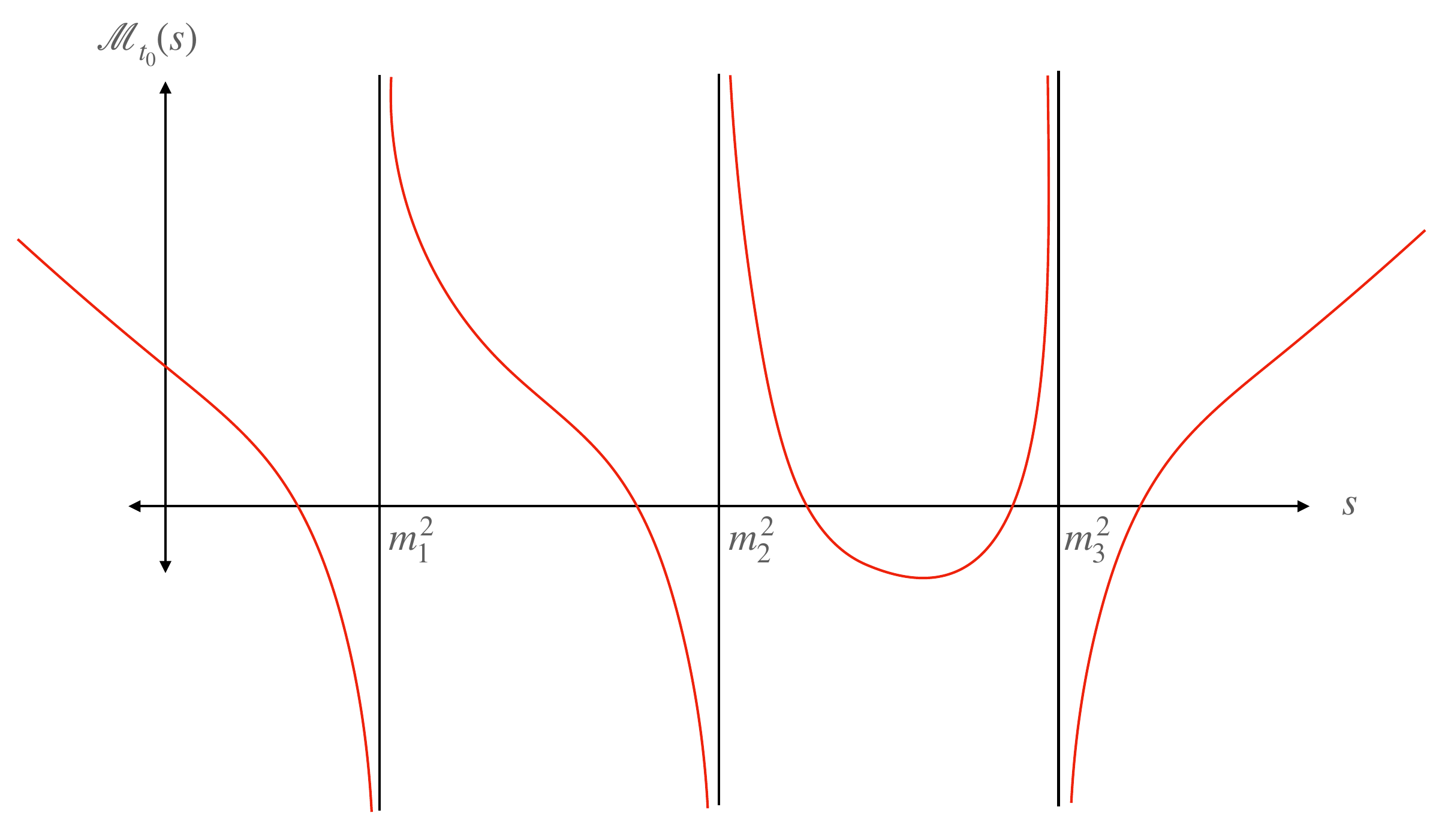}
    \caption{The S-matrix $\CM_{t_0}(s)$ has a single zero between poles $m_1^2$ and $m_2^2$ and hence the residue at both these poles is positive. The number of zeros between $m_2^2$ and $m_3^2$ is two and hence the residue at $m_3^2$ is negative. This shows that the number of zeros between two consecutive poles must be odd for the residues to be positive at all the poles.}
    \label{residue-sign}
\end{figure}

The s-zeros $f_i(t)$ are generally multi-valued functions of $t$, in fact they can be grouped such that $f_{i}(t)$ in a given group are precisely the multiple values of a multi-valued function i.e. values taken by a multi-valued function on different sheets. Two or more $s$-zeros may collide on the $s$ plane as $t$ hits a branch point. 
\begin{framed}
\begin{remark}
    All complex $s$-zeros of $\CM_t(s)$ come in complex conjugate pairs.
\end{remark}
\end{framed}
\noindent
This is due to the reality property $\CM(s,t)=\CM^*(s^*,t^*)$. In particular, this means that as we vary $t$ on the real line, if complex $s$-zeroes ever touch the real line, they must do so in complex conjugate pairs. Similarly, if they do leave the real line they must also do so in complex conjugate pairs. As a result an isolated real $s$-zero remains real at least until it encounters another real $s$-zero.
\begin{framed}
    \begin{remark}\label{oddzeros}
        For $t>4m_\phi^2$, there have to be an odd number of $s$-zeros between any two neighboring $s$-poles.
    \end{remark}
    \end{framed}
    \noindent
This is because for $t>4m_\phi^2$, the residue at any $s$-pole must be positive. Having an even number of zeros would flip the sign of the residue as indicated in figure \ref{residue-sign}.
\begin{framed}
\begin{remark}\label{poly}
    As $t$ approaches $m_i^2$, all $s$-poles are canceled by $s$-zeros such that there are only finitely many $s$-zeros left. The number of leftover zeros is the degree of the residue polynomial at $t=m_i^2$. Moreover, all real zeros of $\CM_{m_i^2}$ must be less than $4m_\phi^2$.  
\end{remark}
\end{framed}
\noindent
This simply follows from the fact that the residue at $t=m_i^2$ is a polynomial in $s$ and that this polynomial is strictly positive for $s>4m_\phi^2$. Having a real zero that is greater than $4m_\phi^2$ would result in the residue polynomial changing sign for $s>4m_\phi^2$, leading to a contradiction. In particular, this also means that the $s$-zero that cancels an $s$-pole must be an isolated zero moving along the real line.
\begin{framed}
\begin{remark}\label{onlyatpoles}
    If for some value $t=t_0 > 4m_\phi^2$, one of the s-zeros coincides with s-pole then $t_0=m_i^2$. This also means that if one of the $s$-pole is cancelled by an $s$-zero then all the $s$-poles must be cancelled by $s$-zeros.
\end{remark}
\end{framed}
\noindent
Let's say the $s$-zero in question is cancelling the $s$-pole at $m_j^2$ and that $t_0$ is not one of the poles but a regular point. The residue at $s=m_j^2$ is zero at $t=t_0$ because it is canceled by one of the $s$-zeros. However, we expect the residue to be strictly positive for $t>4m_\phi^2$ and as $t_0$ is greater than $4m_\phi^2$, it must be that $t=t_0$ is a singularity. 

Before we discuss the movement of $s$-zeros, it is convenient to develop some terminology. Let us denote by $S_i$ the set of the zero functions that cancel all the $s$-poles at $t=m_i^2$. It is also useful to define $\tilde S_i$ to be $S_i$ minus the zero function that is smallest at $t=m_i^2$. 
We now describe how the zeros of set $S_i$ move along the real line as $t$ moves along the real line in the neighborhood of $m_i^2$. 
\begin{framed}
\begin{prop}\label{leftmove}
    As $t$ crosses $m_i^2$ towards right, the zero functions in the set $S_i$ must cross the $s$-poles towards left. 
\end{prop}
\end{framed}
\begin{proof}
Let us focus on a particular zero function $f(t)\in S_i$ that cancels the $s$-pole, say $s=m_j^2$  when $t=m_i^2$ i.e. $f(m_i^2)=m_j^2$. We isolate this zero of the S-matrix and its poles at $t=m_i^2$ and $s=m_j^2$ to define $\CC(s,t)$.
\begin{align}
    \CM(s,t)=\frac{s-f(t)}{(s-m_j^2)(t-m_i^2)} \CC(s,t).
\end{align}
We fist compute the residue of $\CM(s,t)$ at $t=m_i^2$ and evaluate it on $s=m_j^2$.
\begin{align}
    {\rm Res}_{t=m_i^2} \CM(s,t) &= \frac{s-m_j^2}{s-m_j^2} \CC(s,m_i^2)\notag\\
    {\rm Res}_{t=m_i^2} \CM(m_j^2,t)&= \CC(m_j^2,m_i^2).
\end{align}
In the first line, we have used the property $f(m_i^2)=m_j^2$. As the residue polynomial must be positive for any value of $s>4m_\phi^2$, it must be that $\CC(m_j^2,m_i^2)>0$.
Now we compute the residue of $\CM(s,t)$ at $s=m_j^2$ and evaluate it on $t=m_i^2$.
\begin{align}
    {\rm Res}_{s=m_j^2} \CM(s,t) &= \frac{m_j^2-f(t)}{t-m_i^2} \CC(m_j^2,t)\notag\\
    {\rm Res}_{s=m_j^2} \CM(s,m_i^2)&= -f'(m_i^2)\CC(m_j^2,m_i^2).
\end{align}
In the second line, we have evaluated the limit of the residue as $t\to m_i^2$. By same argument as before, this quantity must also be positive. This implies $f'(m_i^2)<0$. Note that this is a strict inequality. It proves that as $t$ crosses $m_i^2$, the zero functions of the set $S_i$ must cross the $s$-poles towards right. 
\end{proof}

\subsection*{A picture for movement of zeros}
Let us summarize the remarks made so far about $s$-zeros. 
At $t=m_i^2$, $s$-zeros coincide with all the $s$-poles with a finitely many $s$-zeros leftover. The number of leftover zeros is the degree of the residue polynomial at $t=m_i^2$ or equivalently, the maximum spin of the particle with mass $m_i^2$. Among the leftover zeros, the real ones must be less than $4m_\phi^2$.

Proposition \ref{leftmove} leads to the following picture about the movement of $s$-zeros. 
As $t$ crosses $m_i^2$ towards right, the zeros of set $S_i$ crosses the $s$-poles towards left, for all $i$ at least locally in $t$ i.e. in some  neighborhood of $t=m_i^2$. As $t$ reaches the next pole $m_{i+1}^2$, the $s$-poles are again cancelled by $s$-zeros of set $S_{i+1}$ traveling to the left.  This picture for the movement of zeros was tacitly assumed in \cite{Caron-Huot:2016icg}. Proposition \ref{leftmove} shows that this movement of zeros to the left as $t$ moves to the right is simply a consequence of positivity of the S-matrix. 

\begin{figure}
    \centering
    \includegraphics[width=1.04\textwidth]{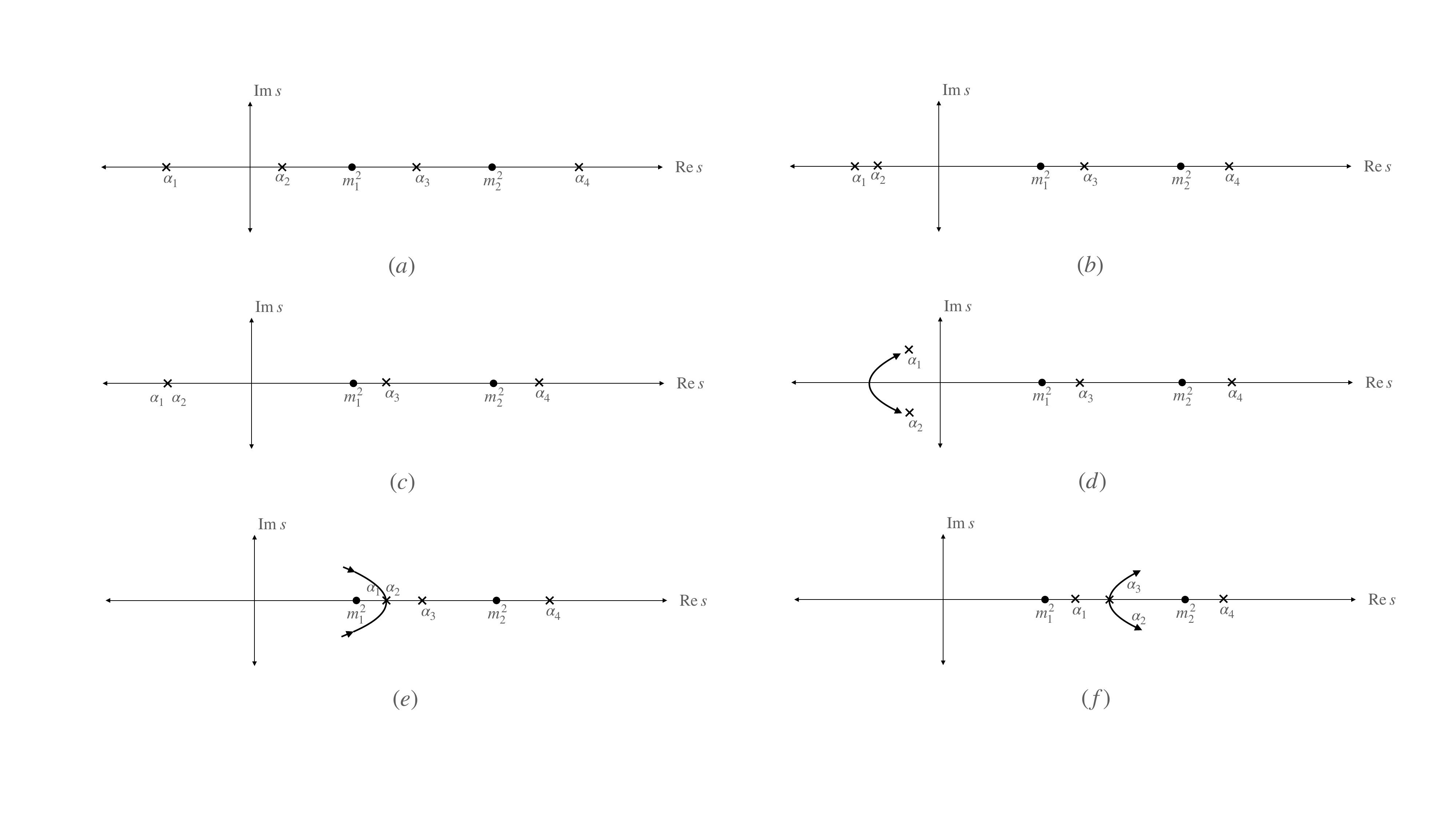} 
   \caption{Recombination of three zeros (crosses) in the complex plane as \( t \) increases, from panels (a) to (f), with poles as dots}
    \label{fig:recombination}
\end{figure}

A priori there is no relation between the sets $S_i$ and $S_{i+1}$ because as $t$ moves from $m_{i}^2$ to $m_{i+1}^2$, some of the complex zeros may come on the real line in complex conjugate pairs, possibly even from infinity, and recombine with the zeros from the set $S_i$ and go off in the complex plane leaving behind  their ``partner'' on the real line. 
This partner may in turn undergo a similar recombination and so on. 
The zeros that are eventually remaining on the real line $> m_1^2$ form the set $S_{i+1}$. In this way, the set $S_{i}$ and $S_{i+1}$ could be completely different due to recombinations. This possibility was not entertained in \cite{Caron-Huot:2016icg}. An example of this kind of recombination is shown in the figure \ref{fig:recombination}. 

\subsection{Observations about {\tt NR} S-Matrices}
If the S-matrix is {\tt NR}, then the zeros of the set ${\tilde S}_i$ do not undergo recombination by definition. This directly implies the following:

\begin{framed}
    \begin{prop}
        If the S-matrix is {\tt NR}, then $S_{i+1} = \tilde S_i$.
    \end{prop}
\end{framed}
\noindent 
This result establishes a train-like picture for the movement of zeros in an {\tt NR} S-matrix. As $t$ moves to the right, the ``train'' of $s$-zeros moves to the left, stopping at $s$-poles precisely when $t$ coincides with a pole. This behavior allows us to assign natural labels to the zero functions of the {\tt NR} S-matrix. Specifically, we label the zero function  that cancels the pole at $s = m_i^2$ when $t = m_1^2$ as $f_i(t)$. Not all zero functions can be labeled this way. The labeled  zero functions are essentially the zeros from the set $S_1$.  For the remaining finitely many zeros we use a different set of labels. We will also extend this zero-labeling scheme to  S-matrices that are not {\tt NR}. 

For {\tt NR} S-matrices, as $t$ varies from $m_1^2$ to $m_j^2$, the zero $f_j(t)$ shifts from $m_j^2$ to $m_1^2$ along the real line. In fact, we can make a stronger statement:
\begin{framed}
    \begin{prop}\label{trainlike}
        For an {\tt NR} S-matrix, $f_j(m_i^2) = m_{j-i+1}^2$ for $j - i \geq 0$.
    \end{prop}
\end{framed}
\noindent
This proposition is a straightforward consequence of the train-like motion of the zeros of an {\tt NR} S-matrix.

Some more results can be established for {\tt NR} S-matrix. 
\begin{framed}
    \begin{remark}\label{self-inverse}
        For an {\tt NR} S-matrix, $f_i(f_i(t)) = t$ for $4m_\phi^2 <t< m_i^2$.
    \end{remark}
\end{framed}
\noindent 
Remark \ref{fjisinverse} states that $f_j(f_i(t)) = t$ for some $j$. Proposition \ref{trainlike} further asserts that $f_i(m_k^2) = m_{i-k+1}^2$. By evaluating $f_i$ on both sides, we find:
\[
f_i(f_i(m_k^2)) = f_i(m_{i-k+1}^2) = m_k^2.
\]
This observation implies that $f_j$ in Remark \ref{fjisinverse} must indeed be $f_i$. If this were not the case, we would have two distinct zero functions, $f_i$ and $f_j$, that both evaluate to $m_k^2$ at $t = f_i(m_k^2)$. Such a scenario is impossible, as poles must be canceled by isolated zeros. Outside the given range of $t$, the zero function $f_i(t)$ can have a branch-cut so the above argument ceases to be valid.

\begin{framed}
    \begin{lemma}\label{monotonic-train}
        For an {\tt NR} S-matrix, the zero function $f_i(t)$ is monotonically decreasing from $m_i^2$ to $m_1^2$ as $t$ increases from $m_1^2$ to $m_i^2$.
    \end{lemma}
\end{framed}

\begin{proof}
If $f_i(t)$ were not monotonic in the interval $\CI = (m_1^2, m_i^2)$, then for some $t_0 \in \CI$, $f_i(t)$ would have the same value at $t_0 - \epsilon$ and $t_0 + \epsilon$. Due to the self-inversion property of $f_i(t)$ in the interval $\CI$, this would lead to a branch cut for $f_i(t)$ within $\CI$. However, this contradicts the fact that $f_i(t)$ does not recombine and remains real in the interval $\CI$. 
\end{proof}

\noindent
This lemma will later help us demonstrate that the Regge trajectory $j(t)$ of {\tt NR} S-matrices is monotonically increasing for $t > m_1^2$, as indicated in Theorem \ref{regge-theorem}. 

\subsection{{\tt NIR} and {\tt PIE}}

A train-like movement of zeros can also be established for {\tt NIR} S-matrices. However, in this case, the train-like behavior can only be established for zeros that are sufficiently far away. More precisely:

\begin{framed}
    \begin{lemma}\label{train-ie}
        For an {\tt NIR} S-matrix, there exists an integer $N_{i}$ such that $f_{i+j}(m_i^2) = m_{j+1}^2$ for $j \geq N_{i}$.
    \end{lemma}
\end{framed}

\noindent
We now provide a lemma relating the concept of intrinsic extremality to the non-recombination of zeros.

\begin{framed}
    \begin{lemma}\label{pie-nir}
        A {\tt PIE} S-matrix without accumulating poles is {\tt NIR}.
    \end{lemma}
\end{framed}

\begin{proof}
A priori, it is possible for the zeros of the  S-matrix to undergo an infinite sequence of recombinations as shown in figure \ref{fig:recombination1}, as $t$ moves from say, $m_i^2$ to $m_{i+1}^2$. Due to {\tt PIE} condition, however, the speed of each zero is bounded as $t$ makes this transition. As a result, the cascade of recombinations shown in figure \ref{fig:recombination1} only travels a finite distance resulting in the S-matrix being {\tt NIR} i.e. it can not undergo infinite recombination.  
\begin{figure}[ht]
    \centering
    \includegraphics[width=0.8\textwidth]{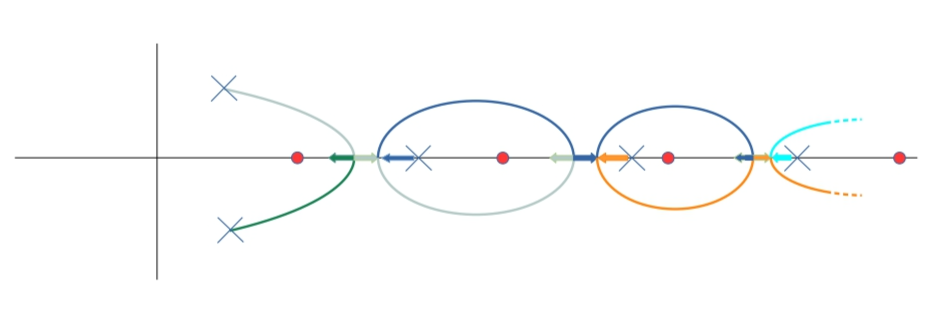} 
    \caption{A possible Infinite recombination of zeros depicted with trajectories in different colors.}
    \label{fig:recombination1}
\end{figure}
\end{proof}
\noindent 
Note that this reasoning applies only in the absence of accumulation points. When the poles have accumulation point, infinite recombinations can occur even if the zeros travel a finite distance, as long as they move towards the accumulation point.  For S-matrices with accumulation points, we have to impose {\tt NIR} condition separately to obtain interesting results like lemma \ref{NIR-degree-increase} and theorem \ref{main-theorem2}.

The properties, {\tt NIR} and {\tt PIE} in combination, are quite powerful as they constrain the degree of the residue  polynomial. This is as follows. 
As $t$ moves from $m_i^2$ to $m_{i+1}^2$ along the real line, there exists a circle $C$ of large radius $R$, centered at the origin, such that zeros not belonging to $\tilde S_i$ do not cross this circle. Consequently, the zeros of $\tilde S_i$ that lie outside this circle do not recombine. Any zeros of $\tilde S_i$ that could potentially recombine are confined within $C$, and these are finitely many. As a result, the number of excess zeros i.e. zeros that are not included in $S_i$, increase precisely by $1$ as $t$ moves from $m_i^2$ to $m_{i+1}^2$.
Let the degree of the residue polynomial at $t = m_i^2$ be $d_i$. It is same  as the number of excess zeros. Hence we have,
\begin{framed}
    \begin{lemma}\label{degree-increase}
        For {\tt PIE} S-matrices without accumulating poles, $d_{i+1} = d_i + 1$ for all $i$.
    \end{lemma}
\end{framed}
\noindent More generally,
\begin{framed}
    \begin{lemma}\label{NIR-degree-increase}
        For {\tt PIE} and {\tt NIR} S-matrices, $d_{i+1} = d_i + 1$ for all $i$.
    \end{lemma}
\end{framed}

\section{Intrinsically extremal S-matrices}\label{intrinsicS}
In section \ref{zero-property} we analyzed local properties of the zeros of the S-matrix. In this section we focus on intrinsic extremality, show how it it parallels the notion of extremality and also how it constrains global properties of zeros.  

Before we go on to prove theorem \ref{ie-finite} for \ie\ S-matrices, let gain some intuition about why it could be true with a simple example. This example also illustrates why the property of intrinsic extremality is morally equivalent to extremality in perhaps the simplest way. Consider the S-matrix $\CM_{m}^{\rm scalar}(s,t)$ corresponding to a scalar exchange.
\begin{align}
    \CM_{m}^{\rm scalar}(s,t)\equiv \frac{1}{s-m^2}+\frac{1}{t-m^2}.
\end{align}
Consider the S-matrix $\CM\equiv \CM_{m_1}^{\rm scalar}+\CM_{m_2}^{\rm scalar}$. As it is a convex sum of S-matrices, it is not an extremal S-matrix. Now we will show that it is also not intrinsically extremal.
Take $t=\frac{m_1^2+m_2^2}{2}+\epsilon$. The $t$-dependent part of the S-matrix has a zero at $\epsilon=0$. The full S-matrix has a zero if we further take either $s=\frac{m_1^2+m_2^2}{2}$ or $s=\infty$. Let's look at the later case. When $\epsilon$ is taken to be small but nonzero the zero of the full S-matrix scales as $s=1/\epsilon$. This means the S-matrix $\CM$ is not \ie.

Recall that the general positive S-matrices have a considerable  multiplicative freedom outlined in remark \ref{const-zero} in the absence of the (classical Regge growth) condition. After imposing the \ie\ condition this freedom reduces to some extent.
\begin{framed}
\begin{remark}\label{no-const-zero}
    If $\CM(s,t)$ is an \ie\ S-matrix then so is $\CB(s,t) \CM(s,t)$ where $\CB(s,t)$ obeys the conditions outlined in remark \ref{const-zero} and it has a fixed number of finitely many $s$-zeros for any value of $t$. 
\end{remark} 
\end{framed}
\noindent The reason $\CB(s,t)$ needs to have finitely many zeros is that $\CB(m_i^2,t)$ is a polynomial. Due to \ie\ property, we do not expect for the number of zeros to change as we change $t$. Hence the function $\CB(s,t)$ has a fixed number of $s$-zeros. In particular, this removes the freedom of multiplying by $\prod_{\alpha\in S}(s-\alpha)(t-\alpha)$ because at $t=\alpha$, the entire polynomial vanishes for all values of $s$. In particular, an \ie\ S-matrix can not have a constant zero i.e. a zero in $s$ that is independent in $t$ (and of course, vice versa). The freedom of multiplying by $\exp(C(s)C(t))$ where $C(s)$ is an entire function with zeros at $m_i^2$ still exists. 

Let us begin the proof of theorem \ref{ie-finite}. 
\begin{shaded*}
\iefinite*
\end{shaded*}
\begin{proof}
    Let us first show that an \ie\ S-matrix with a single pole does not have an entire function remainder. If it does then it can be written as 
    \begin{align}
        \CM(s,t)=\CM_1(s,t)+\CB(s,t).
    \end{align}
    Here $\CM_1(s,t)$ is an S-matrix with a single pole, say at $m_1^2$,  with no remainder and $\CB(s,t)$ is an entire function. By Liouville's theorem, the function $\CB_t(s)$ must be either unbounded or constant. Let us consider the case when it is unbounded. As we take $t\to m_1^2$, $\CM(s,t)\to \infty$. This value can be cancelled by $\CB_t(s)$ as we take $s\to \infty$ because any complex number can be realized as the value of entire function. This shows that as $t\to m_1^2$, one of the $s$-zeros shoots off to infinity. Hence such S-matrix is not \ie.  

    Now we will show that there can not be an \ie\ S-matrix with finitely many but more than one pole. Let us assume the contrary. Let the highest pole be at $m_k^2$. 
    As $t\to m_1^2$, all the $s$-poles are cancelled by $s$-zeros. Let us focus on the cancellation at $s=m_k^2$. Thanks to remark \ref{poly}, when this cancellation occurs, there can not be any leftover $s$-zeros on the real axis beyond $s=m_k^2$. As $t$ moves to the next pole $m_2^2$, the pole at $s=m_k^2$ must again be cancelled by an $s$-zero coming from right. As there were no real zeros in that region at $t=m_1^2$, for this to happen, some complex zeros must come on the real line. This must occur in pairs. Out of this even number of zeros, one cancels the pole at $s=m_k^2$ as $t\to m_2^2$. This means, at this value to $t$, there must be odd number zeros leftover on the real axis beyond $s=m_k^2$. This contradicts the remark \ref{poly}. 
\end{proof}

\subsection{\ie\ S-matrices with infinitely many zeros}
In this section we will prove the theorem \ref{connection} for \ie\ S-matrices with infinitely many poles using the lemma \ref{degree-increase}. We will then prove the theorem \ref{regge-theorem}.

\begin{shaded*}
\connection*
\end{shaded*}
\begin{proof}
    Let us index the poles of $\CM= \sum_{i=1}^n p_i \CM_i$, i.e. elements of $K$ by $i$ in increasing order. At every pole $m_i^2\in K$, we will denote the degree of the residue polynomial of $\CM$ as $d_i$. At $m_i^2\in K$ we also define the degree of the residue polynomial of the S-matrix $\CM_\alpha$ as $d_i^{(\alpha)}$. If some $\CM_\alpha$ does not have a pole at $m_i^2$ then $d_i^{(\alpha)}$ is taken to be undefined. Note that $d_i=\max\{d_i^{(1)},\ldots d_i^{(n)}\}$.

    Let us see how $d_i^{(\alpha)}$ behaves with $i$ for some $\alpha$. Thanks to lemma \ref{degree-increase}, $d_{i}^{(\alpha)}$ increases by $1$ with $i$ for all $i$ where it is defined. It leads to the following graph. 
    \begin{figure}[h]
        \centering
        \includegraphics[scale=0.30]{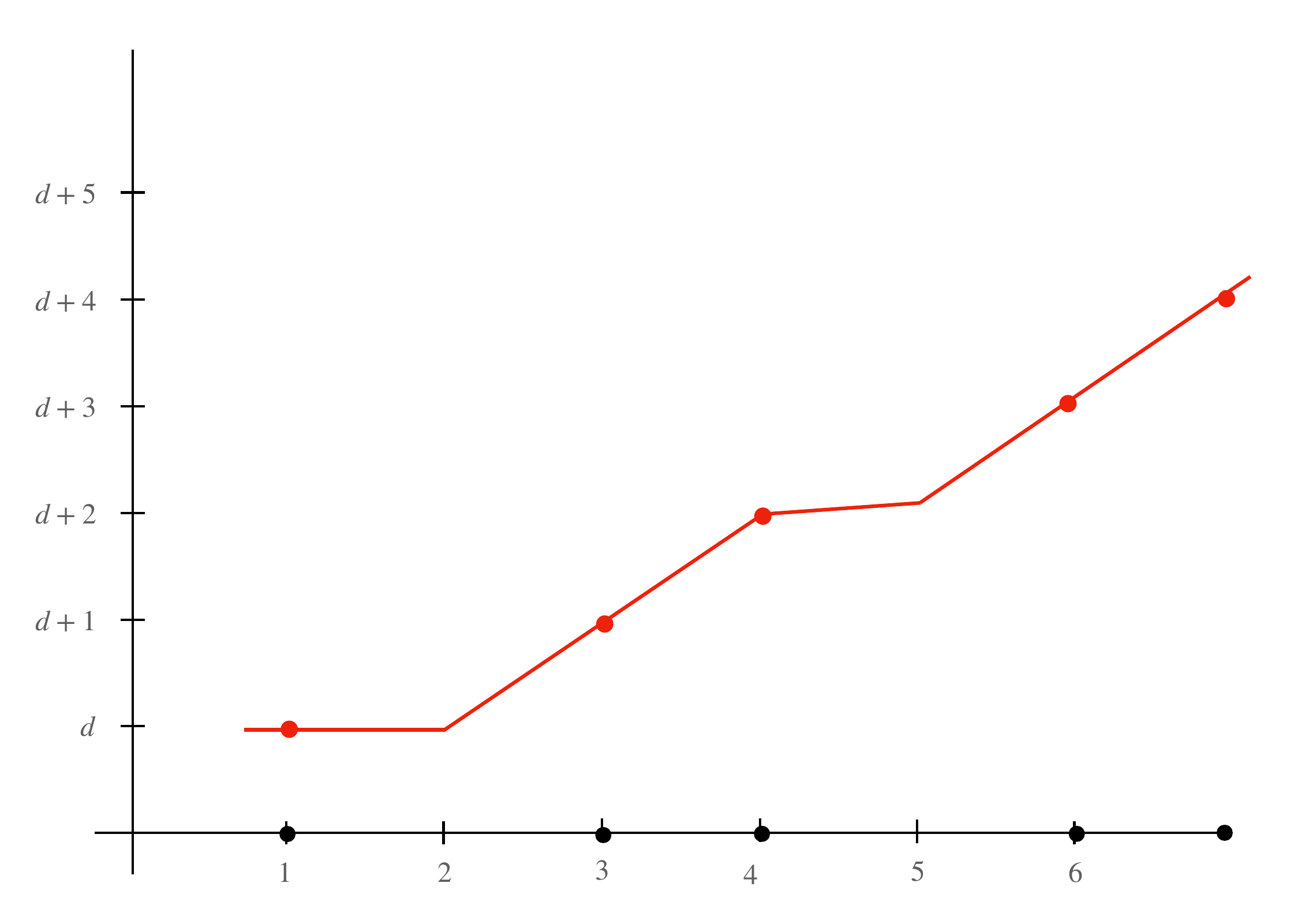}
        \caption{Degrees $d_i^{(\alpha)}$ of the residue polynomial of the S-matrix $\CM_\alpha$. It shows monotonic growth by $1$ at every pole of  $\CM_\alpha$. On $x$-axis we have schematically marked poles of all the S-matrices i.e. the poles that are in $K$, with poles in $K_\alpha$ highlighted.}
    \end{figure}
    Because of the condition $\min\{K_\alpha\} < \min\{K\setminus K_\alpha\}$, there exists at least one pole $m_i^2$ such that $i<j$ for any $j$ such that $d_j^{(\alpha)}$ is not defined. Because the degree of the residue polynomial of $\CM$ obeys $d_i=\max\{d_i^{(1)},\ldots d_i^{(n)}\}$, it can be computed by superimposing the graphs of $d_i^{(\alpha)}$ for all $\alpha$ and finding its upper envelope. If $\CM$ were \ie\ then this envelope must be a graph that increases by $1$ at every $i$. By inspection, it is clear that graphs of $d_i^{(\alpha)}$ can not superimpose to yield such an envelope.
    
    Note that for this argument to work, the graph of $d_i^{(\alpha)}$ must have the zero before the ``skipped''  pole. This is simply the condition mentioned in the theorem. 
\end{proof}

\noindent Now we move to the proof of theorem \ref{regge-theorem}
\begin{shaded*}
\reggetheorem*
\end{shaded*}
\begin{proof}
    Let us take $t\in (m_n^2, m_{n+1}^2)$. At this value of $t$, the zeros of $\CM$ are naturally divided into two sets. The set of zeros that would cancel $s$-poles at $t=m_{n+1}$ i.e. $S_{n+1}$ and the rest $R$. The zeros in $S_{n+1}$ are real and there is precisely one of them between any two poles. Let us label the zero appearing right after the pole $m_i^2$ as $\alpha_i$. The zeros of the set $R$ are denoted as $\beta_a$. The degree of the residue polynomial at $t=m_n^2$ is $j(m_n^2)$. Due to lemma \ref{degree-increase}, this makes the degree of the residue polynomial at $t=m_{n+1}^2$ to be $j(m_{n+1}^2)=j(m_n^2)+1$. The zeros in $R$ precisely correspond to the zeros of the residue polynomial at $t=m_{n+1}^2$. This means $|R|=j(m_n^2)+1$. 
    We would like to show that for  $t\in (m_n^2, m_{n+1}^2)$, $j(t)\in (j(m_n^2),j(m_n^2)+1)$. 

    Let us define,
    \begin{align}
        f(s,t)=\frac{\log(\CM(s,t))}{s-\sigma}
    \end{align}
    which has the property that $f(s,t)\to 0 $ for $|s|\to \infty$. We consider the contour integral
    \begin{align}
        \frac{1}{2\pi i}\oint_{C_{s_0}} \frac{f(s,t)}{s-s_0} = f(s_0,t),
    \end{align}
    where the contour is taken to be a small circle around $s_0$. We deform this contour to surround other features in the $s$ plane along with the contour $C_{\infty}$ at infinity. The contribution from $C_{\infty}$ vanishes as $f(s,t)\to 0 $ there. 
 
    The singular features of $f(s,t)$ include a pole at $\sigma$, semi-infinite branch cuts $C_{\beta_a}$ starting from zeros $\beta_a$ $\in R$ of $\CM(s,t)$ that we orient towards $-\infty$ and the branch cuts starting from zeros in $S_{n+1}$ that end on the poles that are immediately to their left. These branch cuts consists of intervals $[m_i^2,\alpha_i]\equiv C_{\alpha_i}$ for all $i$. A picture  of these singular features is shown in figure \ref{sing-features}. 
    \begin{figure}[h]
        \centering
        \includegraphics[scale=0.3]{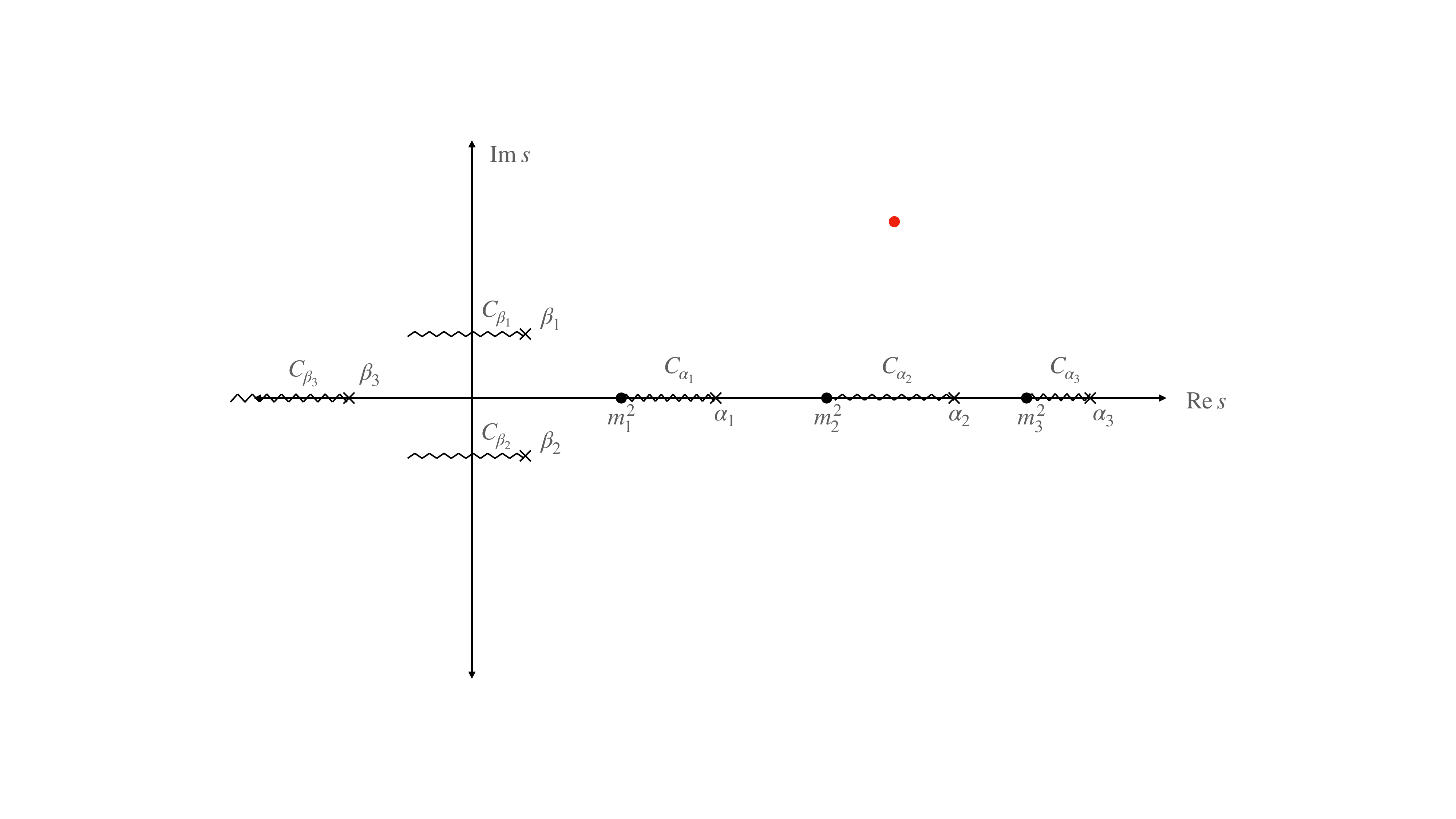}
        \caption{Singular features of $f(s,t)$. The branch cuts $C_{\beta_a}$ and $C_{\alpha_i}$ are shown. The pole at point $\sigma$ is denoted by a red dot.}
        \label{sing-features}
    \end{figure}
    Deforming the contour $C_{s_0}$ to enclose all the singular features we get,
    \begin{align}
        f(s_0,t)&=\frac{\log \CM(s_0,t)}{s_0-\sigma} =-\frac{1}{2\pi i}\oint_{C_{\sigma}+\sum C_{\beta_a}+\sum C_{\alpha_i}} \frac{f(s,t)}{s-s_0} \\
        \log \CM(s_0,t)&= \log \CM(\sigma,t)+ \sum_a \Big(\log(s_0-\beta_a)-\log(\sigma-\beta_a)\Big) \notag \\
        &+ \sum_i \Big(\log\frac{s_0-\alpha_i}{s_0-m_i^2}-\log\frac{\sigma-\alpha_i}{\sigma-m_i^2}\Big).\notag
    \end{align}
    Here we have used the fact that the discontinuity around each cut is $2\pi i $. This equation implies
    \begin{align}\label{s-indep}
        \log \CM(s_0,t) -\sum_a \log(s_0-\beta_a) - \sum_i \Big(\log\frac{s_0-\alpha_i}{s_0-m_i^2}-\log\frac{\alpha_i}{m_i^2}\Big) = (s_0\to \sigma ).
    \end{align}
    The role of the term $\log (\alpha_i/m_i^2)$ is to make the sum over $i$ converge. Let us show this explicitly,
    \begin{align}
        \sum_i \Big(\log\frac{s_0-\alpha_i}{s_0-m_i^2}-\log\frac{\alpha_i}{m_i^2}\Big) &= \sum_i \log\frac{1-s_0/\alpha_i}{1-s_0/m_i^2}\notag\\
        &\xrightarrow{i\to \infty} s_0\sum_i \Big(\frac{1}{m_i^2}-\frac{1}{\alpha_i}\Big)< \sum_i \Big(\frac{1}{m_i^2}-\frac{1}{m_{i+1}^2}\Big).
    \end{align}
    This sum telescopes and gives a finite answer. Getting back to equation \eqref{s-indep}, it tells us that the left hand side doesn't depend on $s_0$ because $s_0$ can be changed to any $\sigma$ to get the same quantity. This means
    \begin{align}
        \log \CM(s_0,t) &\xrightarrow{s_0\to \infty} (j_{m_{n+1}^2}+ c) \log s_0 + A(t)\notag\\
        {\rm where} \quad c&= \lim_{s_0\to \infty }\frac{1}{\log s_0} \sum_i \Big(\log\frac{s_0-\alpha_i}{s_0-m_i^2}-\log\frac{\alpha_i}{m_i^2}\Big)
    \end{align}
Here $A(t)$ doesn't depend on $s_0$. Now we will show $-1<c<0$. We will take $s$ to be large negative for convenience but the argument can be made for any large $|s|$ with fixed phase. It is clear that the quantity in the bracket is negative and monotonically decreasing with respect to each $\alpha_i$. To show that $c>-1$, we take the maximum value of $\alpha_i$ namely $\alpha_i=m_{i+1}^2$. Due to monotonicity, this achieves the minimum. With this choice the sum telescopes, we get $c=-1$. 

For S-matrices that are {\tt NR}, lemma \ref{monotonic-train} holds. Hence the $s$-zeros are monotonically decreasing  with $t$. This means that above $\alpha_i$'s are monotonically decreasing and so $j(t)$ is monotonic for $t>m_1^2$. 
\end{proof}
\noindent
It is easy to see why $|s|^{j(t)}$ behavior of $\CM(s,t)$ in the Regge limit is necessary for this theorem to hold. We can multiply $\CM(s,t)$ by $\exp (\CC(s)\CC(t))$ where $\CC(s)$ is an entire function that has zeros at $m_i^2$, and obtain an \ie\ S-matrix that has the same residue polynomial but violates the boundedness used in theorem \ref{regge-theorem}. It is clear that the new S-matrix, even though it is \ie\, does not behave as $|\CM(s,t)|\sim |s|^{j(t)}$ for any $j(t)$. 

\section{Enhanced crossing symmetry}\label{enhanced}
In this section, we will construct a canonical form for the S-matrices with infinitely many poles. Using the canonical form, we will define the notion of enhanced crossing symmetry and then explore its consequences. Before we go on to do so, let us take a brief detour into complex analysis to discuss Weierstrass factorization of entire functions. 

A polynomial $P(z)$ of degree $d$ can always be written as a product of $d$  factors as $P(z)=(z-z_1)\ldots (z-z_d)$, where $z_i$ are its zeros. This is not true as is for an entire function with infinitely many zeros. This is because infinite product of factors of the type $(z-z_i)$ may not be convergent. However, this problem can be fixed by attaching a simple ``nowhere vanishing entire function'' piece to each of these factors. This is best illustrated with the help of an example. Consider expressing the entire function $1/\Gamma(1-z)$ as a product of its factors. It has zeros at $z=1,2, \ldots$. So the first guess would be
\begin{align}
    \frac{1}{\Gamma(1-z)}\stackrel{?}{=}\prod_{n=1}^{\infty}(1-\frac{z}{n})=\exp(-\sum_{n=1}^{\infty} \sum_{k=1}^\infty\frac{1}{k} \frac{z^k}{n^{k}}). 
\end{align}
On the right hand side we have expanded $\log(1-z/n)$ in a power series. The $n$ sum becomes $\sum_n n^{-k}$. This sum is convergent for $k\geq 2$ but not for $k=1$. In order to cancel this divergence, we need to remove the $k=1$ term. This is done if the above infinite product is replaced by
\begin{align}
    \prod_{n=1}^\infty (1-\frac{z}{n}) e^{\frac{z}{n}}. 
\end{align} 
This is a convergent product. This is equal to $1/\Gamma(1-z)$ up to a nowhere vanishing entire function.  

These ideas are formalized into the so-called Weierstrass factorization theorem. It says that every entire function $f(z)$ can be written as\footnote{If there is a zero of $f$ at $z=0$ then the right hand side needs to be multiplied by $z^k$ where $k$ is the order of the zero.}
\begin{align}
    f(z)=\prod_{n=1}^\infty (1-\frac{z}{z_n}) E_p(\frac{z}{z_n}), \qquad {\rm where}\quad E_{p}(x)=\exp(\sum_{k=1}^{p} \frac{x^k}{k}).
\end{align}
Note that this doesn't specify the entire function uniquely given its zeros. This is expected since an entire function can always be multiplied by a nowhere vanishing entire function, such as $e^{g(z)}$ where $g(z)$ is an entire function, without affecting its zeros. 

\subsection{Possible singularities of the zero functions}\label{f-sing}
To define the canonical form of the S-matrix we need to assume that 
\begin{itemize}
    \item Assumption 1: all singularities of $f_n(t)$ are isolated.
    \item Assumption 2: $f_n(t)$ does not have a branch point of infinite order.
\end{itemize}
\noindent 
All of our theorems, with the exception of theorem \ref{main-theorem2}, are for \ie\ S-matrices i.e. they assume that the zero functions $f'_n(t)$ do not have any singularity. So the Assumption 1 has a bearing only on the proof of theorem \ref{main-theorem2} which applies for the S-matrix that is partially \ie\ and hence leaves open the possibility that the zero functions may have singularities somewhere in the complex plane. It would be good to prove these assumptions for partially \ie\ S-matrices. 
In section \ref{examples}, we will give examples of S-matrices that violate the Assumption 1 as they have a type of non-isolated singularity called a ``natural boundary''. However, none of these examples are partially \ie. 

Let us comment briefly on Assumption 2.  
\begin{framed}
    \begin{remark}
        If $f_n(t)$ has a branch point with infinitely many sheets then there must be a singularity at this branch point.
    \end{remark}
\end{framed}
\noindent
Lets assume that $f_n(t)$ doesn't have a singularity at this branch point $t_0$  but rather takes the value $f_0$. Because this happens all infinitely many sheets,  infinitely many $s$-zeros meet at $f_0$ when $t=t_0$. This makes $t=t_0$ an essential singularity of $\CM_{t_0}(s)$, contradicting the meromorphic property of the S-matrix. On the other hand, if $f_n(t)$ does diverge at $t_0$, e.g. as $\log (t-t_0)$ then infinitely many $s$-zeros go off to infinity at $t=t_0$. This, by itself does not contradict the meromorphic property of the S-matrix. However, we still consider this case somewhat pathological and avoid it with Assumption 2. 

Now we show that 
\begin{framed}
    \begin{remark}\label{no-essential}
        $f_n(t)$ can not have an essential singularity. 
    \end{remark}
\end{framed}
\noindent
We will show that if $f_n(t)$ has an essential singularity then $\CM(s,t)$ also has an essential singularity leading to a contradiction. Let this essential singularity be at $t_0$. Due to Picard's great theorem, in  neighborhood of $t_0$, however small, we can find $f_n$ taking on any given value, say $s_0$, infinitely many times. This means $(s,t)=(s_0,t_\alpha)$ are zeros of the S-matrix, where $t_\alpha$ are infinitely many points in a small neighborhood of $t_0$. As the S-matrix crossing symmetric, $(s,t)=(t_\alpha, s_0)$ are also its zeros. This means that as $t$ is taken to be $s_0$, infinitely many $s$-zeros gather in a neighborhood of $t_0$, however small. This is a hallmark of an essential singularity of the S-matrix at $t=t_0$. This means that if $f_n(t)$ does have a singularity, it must be removable. If the essential singularity is at a branch point with finitely many  $m$ number of sheets then we can use the same argument using a single valued chart $(t-t_0)^m=t'$ in the vicinity of $t_0$ to rule it out. 

The upshot of the above discussion is that $f_n(t)$ only has singularities of the type $(t-t_i)^{-\alpha_i}$ where $\alpha_i$ are positive rational numbers. In particular, we can find a function $A_n(t)=\prod_{i}(t-t_i)^{\alpha_i}$ with $\alpha_i$ rational such that $A_n(t)f_n(t)$ does not have a singularity.

\subsection{Canonical form of the S-matrix}

Let us get back to the discussion at hand. The S-matrix, $M(s,t)$ is a meromorphic function of $s$ and $t$ with constant poles $m_n^2$, both in $s$ and $t$. If we multiply it by an entire function of $s$ with zeros at $m_n^2$ and similarly for $t$, then the resulting function is entire. In other words, 
\begin{align}
    \CM(s,t)=\frac{N(s,t)}{\prod_{n=1}^{\infty}(1-\frac{s}{m_n^2}) C_n(s) \prod_{n=1}^{\infty}(1-\frac{t}{m_n^2}) C_n(t)},
\end{align}
where $N(s,t)$ is an entire function. The functions $C_n(t)$ are entire function factors that make the $\prod_n$ convergent. Let us think of $N(s,t)$ as an entire function of $s$ for a given value of $t$. Then, using the factorization theorem,
\begin{align}
    N(s,t)=\prod_{n=1}^{\infty} (1-\frac{s}{f_n(t)}) B_n(s,t).
\end{align}
Here $f_n(t)$ are the zeros of $\CM_t(s)$ and $B_n(s,t)$ is a nowhere vanishing  entire function of $s$. Now we will introduce the notion of a canonical form of the numerator $N(s,t)$ where we rearrange the product so that it can be written in terms of another nowhere vanishing entire function $\tilde B_n(s,t)$ of $s$ which does not have zeros and singularities in $t$ as well. 

The function $f_n(t)$ could have a zero. But the S-matrix is not singular there because all the singularities of $\CM(s,t)$ have already been accounted for by the denominator. The singularity coming from the factor $(1-s/f_n(t))$ at the zero of $f_n(t)$ must be cancelled by the zero $B_n(s,t)$. We can absorb $1/f_n$ factor in $B_n$ and rewrite the infinite product $N(s,t)$ as 
\begin{align}
    N(s,t)=\prod_{n=1}^{\infty}(s-f_n(t)){\hat B}_n(s,t),\qquad {\hat B}_n(s,t)=-B_n(s,t)/f_n(t).
\end{align}
The function ${\hat B}_n(s,t)$ does not have any zeros in $t$ that are $s$-dependent. Because those would also be zeros of $s$ and they have already been accounted for by the factor $\prod_{n=1}^{\infty}(s-f_n(t))$. But it  could have s-independent zeros in $t$. These are not zeros of the S-matrix as the S-matrix does not have any constant zeros in $t$. So they must be precisely cancelled by the singularity in $f_n(t)$. In section \ref{f-sing}, we have looked at the possible singularities of $f_n(t)$. 
We remove the $s$-independent zeros of ${\tilde B}_n(s,t)$ into a function $A_n(t)$ and rewrite ${\hat B}_n(s,t)=A_n(t){\tilde B}_n(s,t)$. 
As remarked in section \ref{f-sing}, $A_n(t)$ can be chosen to have the form $\prod(t-t_i)^{\alpha_i}$ where the product is over the singularities $t_i$ of $f_n(t)$ and $\alpha_i$ are rational numbers. With this choice of $A_n(t)$, $A_n(t)(s-f_n(t))$ does not have any singularities. If the $f_n(t)$ is non-singular, as for the case of \ie\ S-matrices, then we choose $A_n(t)=1$. 

Can $\tilde B(s,t)$ have singularities in $t$? If the singularity is not constant then it is a singularity in $s$ as well. This contradicts the fact that $B_n(s,t)$ is an entire function of $s$. If the singularity is at a constant, then it must be a singularity of the full S-matrix as it can not be canceled by the zero of the $(s-f_n(t))$. Such singularities are already accounted for by the denominator. All in all, we have rearranged the numerator into a form
\begin{align}\label{best-form}
    N(s,t)=\prod_{n=1}^{\infty}(s-f_n(t))A_n(t){\tilde B}_n(s,t).
\end{align}
where $\tilde B(s,t)$ does not have zeros or singularities in $s$ as well as in $t$. We call this the canonical form of the S-matrix.  
\setcounter{definition}{5}
\begin{framed}
\begin{definition}[Enhanced crossing symmetry]
    An S-matrix is called enhanced crossing symmetric if its canonical form satisfies
    \begin{align}
       \lim_{N\to \infty} \prod_{n=1}^{N} \frac{\tilde B_n(s,t)}{\tilde B_n(t,s)}\equiv 1.
    \end{align}
    The convergence is uniform in the entire $s$ and $t$ plane. 
\end{definition}
\end{framed}
\noindent
Together with the crossing symmetry of the S-matrix, the condition \ec\ implies
\begin{align}
    \prod_{n=1}^{\infty} \frac{A_n(t)(s-f_n(t))}{A_n(s)(t-f_n(s))}  =1.
\end{align}
Here $\prod_{n=1}^\infty$ is a shorthand for the limit $N\to \infty$ of $\prod_{i=1}^N$. We will use this shorthand in what follows. 

\begin{framed}
\begin{prop}
    The Veneziano amplitude is \ec.
\end{prop}
\end{framed}
\begin{proof}
    The Veneziano amplitude is
\begin{align}
    \CM_V(s,t)=\frac{\Gamma(-1-s)\Gamma(-1-t)}{\Gamma(-2-s-t)}.
\end{align}
In this case, $m_n^2= n-2$ and $f_n(t)=-t+n-3$ for $n=1,2,\ldots$. The numerator takes the form
\begin{align}
    N(s,t)=\prod_{n=1}^{\infty} (1-\frac{s}{-t+n-3}) B_n(s,t),\qquad B_n(s,t)=  (1 - \frac{t}{n - 3}) e^{\frac{s + t}{n - 3}}
\end{align}
As expected $B_n(s,t)$ is a nowhere vanishing entire function of $s$ but it does have zeros in $t$. These zeros precisely cancel the singularity coming from the zero of $f_n(t)$. Cancelling these zeros with singularity, we get the expression
\begin{align}
    N(s,t)= \prod_{n=1}^\infty(s+t-n+3) {\tilde B}_n(s,t), \qquad {\tilde B}_n(s,t)= -\frac{1}{n - 3} e^{\frac{s + t}{n - 3}}.
\end{align}
Observe that each of the two factors inside the product are crossing symmetric separately. In particular
\begin{align}
    \prod_{n=1}^{\infty}\frac{s-f_n(t)}{t-f_n(s)} =  \prod_{n=1}^{\infty}\frac{s+t-n+3}{t+s-n+3} = 1,\qquad{\rm and}\qquad  \prod_{n=1}^{\infty}\frac{{\tilde B}_n(s,t)}{{\tilde B}_n(t,s)}=1.
\end{align}
We extract this property of the Veneziano amplitude into the definition of enhanced crossing symmetry (\ec). 
\end{proof}
\begin{framed}
\begin{prop}
    The Coon amplitude is \ec.
\end{prop}
\end{framed}
\begin{proof}
    The Coon amplitude is given by
    \begin{align}
        \CM_C(s,t) &\propto \prod_{n=0}^{\infty} \frac{\sigma \tau - q^n}{(\sigma - q^n)(\tau - q^n)} \qquad \text{where} \notag \\
        \sigma &= 1 - (s - m^2)(1 - q), \qquad \tau = 1 - (t - m^2)(1 - q),
    \end{align}
    up to some proportionality constant, which is not crucial here.
    
    We can now express this in the standard form as
    \[
    N(s,t) = \prod_{n=1}^{\infty} \left( 1 - \frac{(s - m^2)(1 - q)}{1 - q^n} \Big/ \left(1 - \frac{(1 - q)(t - m^2)}{1 - q^n} \right) \right) B_n(s,t),
    \]
    where
    \[
    B_n(s,t) = \frac{1 - \frac{q^n}{1 - (t - m^2)(1 - q)}}{1 - (s - m^2)(1 - q)}.
    \]
    This can be rewritten in the standard form as
    \[
    N(s,t) = \prod_{n=1}^{\infty} A_n(t) (s - f_n(t)) \tilde{B_n}(s,t),
    \]
    where
    \[
    \tilde{B_n}(s,t) = \frac{1}{(1 - (s - m^2)(1 - q))(1 - (t - m^2)(1 - q))}.
    \]
    Which is crossing symmetric and the functions \(f_n(t)\) and \(A_n(t)\) are given by
    \begin{align}
        f_n(t) &= s - m^2 - \frac{1}{1 - q} + \frac{q^n}{(1 - (t - m^2)(1 - q))(1 - q)}, \notag \\
        A_n(t) &= \left( t - m^2 - \frac{1}{1 - q} \right).
    \end{align}
    \noindent 
    It is now straightforward to verify that \( A_n(t)(s - f_n(t)) \) is crossing-symmetric for each \( n \). This completes the proof that the Coon amplitude is enhanced crossing-symmetric.
\end{proof}

\begin{framed}
\begin{prop}
    An S-matrix with finitely many zeros is \ec.
\end{prop}
\end{framed}
\begin{proof}
Consider the canonical form of the numerator of the S-matrix 
\begin{align}
    N(s,t)=\prod_{i=1}^n A_i(t)(s-f_i(t))B_i(s,t)= F(s,t) B(s,t),
\end{align}
where we have defined $F(s,t)=\prod_{i=1}^n A_i(t)(s-f_i(t))$ and $B(s,t)=\prod_{i=1}^n B_i(s,t)$. Clearly $F(s,t)$ is a polynomial in $s$. We would like to show that $F(s,t)$ is a polynomial in $t$ as well. This, coupled with the fact that the zeros of $F(s,t)$ and $N(s,t)$ are identical, shows that $F(s,t)$ is crossing symmetric. 

As we have taken the product over all $i$'s, $F(s,t)$ is single valued. Also, it does not have any singularities at finite value of $t$ as all such singularities are accounted for by the denominator of the canonical form. To show that it is a polynomial in $t$, we show that as $t\to \infty$ along any direction, $|F(s,t)|\to \infty$. An analytic function obeying this property is necessarily a polynomial.

Remark \ref{no-essential} states that $f_i(t)$ can not have have an essential singularity. The S-matrix with finitely many zeros can not have essential singularity even at $\infty$. This is because, repeating the same argument as in remark \ref{no-essential}, we see that as we approach some value $t_0$, infinitely many zeros go off to infinity. This is not possible because the S-matrix only has finitely many zeros. This means that as $t\to \infty$, $f_i(t)$ either diverges as a power or goes to a constant value. If $f_i(t)$ goes to a constant value $f_0$ as $t\to \infty$, then due to crossing symmetry, it must have a power type singularity at $t=f_0$. All in all, $F(s,t)=\prod_{i=1}^n A_i(t)(s-f_i(t))$ diverges as a power as $t\to \infty$. This implies $F(s,t)$ is a polynomial in $t$.
\end{proof}
\noindent
The condition of \ec\ might seem too abstract and restrictive but it is quite general and comprehensively includes various ansatz for S-matrices (with infinitely many poles) that have been considered in the literature. For example, in \cite{Geiser:2022Gen}, the authors assumed the ansatz
\begin{align}\label{linear-ansatz}
A(s, t) = {\tilde B}(s, t) \frac{1}{st} \prod_{n \geq 1} \frac{1 - A_n (s + t) + B_n st}{(1 - \frac{s}{\lambda_n})(1 - \frac{t}{\lambda_n})}
\end{align}
and showed that an S-matrix with above ansatz must have pole structure coinciding with either Coon or Veneziano Amplitudes. As we will show below, if we generalize the ansatz \eqref{linear-ansatz} to include any any polynomial functions of $(s,t)$ in the numerator then the resulting S-matrix is \ec. This is summarized in the following proposition.
\begin{framed}
    \begin{prop}\label{poly-form}
    Consider an \ie\ S-matrix that admits a canonical form can be written as:
    \begin{align}
        \CM(s,t) = \tilde{B}(s,t) \frac{\prod_{n=1}^\infty P_n(s,t)}{\prod_{n=1}^{\infty} \left( 1 - \frac{s}{m_n^2} \right) C_n(s) \prod_{n=1}^{\infty} \left( 1 - \frac{t}{m_n^2} \right) C_n(t)},
 \end{align}
    where \( P_n(s,t) \) are crossing-symmetric polynomials of bounded degree with bounded spread of zeros. Such an S-matrix is enhanced crossing-symmetric.
    \end{prop}
    \end{framed}
    \begin{proof}
        Consider the product
        \begin{align}
            \prod_{i=1}^{N} \frac{s - f_i(t)}{t - f_i(s)} 
        \end{align}
        for large $N$. We combine the zero functions $f_i(s)$ for $i\leq N$ into zeros of polynomial $P_n(s,t)$. After combining all possible zeros there are a finite number of zeros left-over that are larger than some $R$ for sufficiently large $N$ for $s,t$ in some bounded domain. Let the set of these zeros be $L$. 
        As a result, we have
        \begin{align}
            \prod_{i=1}^{N} \frac{s - f_i(t)}{t - f_i(s)} = \prod_n \frac{P_n(s,t)}{P_n(t,s)}\times \prod_{i\in L} \frac{s - f_i(t)}{t - f_i(s)} = \prod_{i\in L} \frac{s - f_i(t)}{t - f_i(s)} 
        \end{align}
        The second equality follows from the crossing symmetry of $P_n(s,t)$. Now we will show that the product of leftover terms in $L$ goes to $1$ for large $N$. Taking  $(s,t)$ to be, say $(m_1^2,m_2^2)$
        \begin{align}
            \prod_{i=1}^{N} \frac{s - f_i(t)}{t - f_i(s)}|_{(s,t)=(m_1^2,m_2^2)}= \prod_{i\in L} \frac{m_1^2 - f_i(m_2^2)}{m_2^2 - f_i(m_1^2)} = \prod_{i\in L} \frac{m_1^2 - m_{i-1}^2}{m_2^2 - m_{i}^2} =  \prod_{i\in L} \frac{  m_{i-1}^2}{ m_{i}^2}
        \end{align}
        In the last equality, we have used the fact that $m_{i-1}^2 \gg m_1^2$ and $m_i^2 \gg m_2^2$ etc. Using the fact that $m_i^2/m_{i+1}^2$ tends to $1$ large $i$. We get the desired result. This argument is valid for any finite domain of $(s,t)$ near $m_1^2$. 

    \end{proof}
    \noindent
In fact we believe that the notion of \ec\ is very general and conjecture,
\begin{framed}
\begin{conj}
    Any S-matrix is \ec.
\end{conj}
\end{framed}
\noindent 
It would interesting to explore this conjecture further. We will not do so in this paper and treat \ec\ as an independent nontrivial condition.

\subsection{Main theorems}
Now we prove our main theorems, theorem \ref{main-theorem} and \ref{main-theorem2}.
\begin{shaded*}
\maintheorem*
\end{shaded*}
\begin{proof}
    As the S-matrix is \ie\ and \ec, its canonical form obeys $A_n(t)=1$. The \ec\ condition is
    \begin{align}
        \prod_{n=1}^\infty \frac{s-f_n(t)}{t-f_n(s)}=1.
    \end{align}
    Let us denote left hand side evaluated at $(s_0,t_0)$ as $g(s_0,t_0)$. We will consider $g(m_i^2,m_j^2)$ for various values of $i,j$. Each term in the product $\prod_{n=1}^\infty m_i^2 -f_n(m_j^2)$ is non-zero and finite, except for the term corresponding to the  zero function $f(t)$ which cancels the pole at $s=m_i^2$ when $t=m_j^2$. This is true for numerator and denominator possibly with different zero-functions. If we think of $g(m_i^2,m_j^2)$ as a limit, then the ratio of these two vanishing terms gives a finite contribution. It is also non-zero because $g(m_i^2,m_j^2)=1/g(m_j^2,m_i^2)$.
    As a result, 
    \begin{align}
        \prod_{n=n'}^\infty \frac{s-f_n(t)}{t-f_n(s)}|_{(s,t)=(m_i^2,m_j^2)}\equiv g_{n'}(m_i^2,m_j^2)={\rm finite,\,non}{\rm -}{\rm zero},
    \end{align}
    for sufficiently large $n'$. We showed in lemma \ref{train-ie} that for a partially \ie\ S-matrix, in particular for \ie\ S-matrix,  there exists an integer $N_{i}$ such that $f_{i+j}(m_i^{2})=m_{j+1}^2$ for $j\geq N_{i}$. We will consider $n'$ in  $g_{n'}(m_i^2,m_j^2)$ to be greater than the maximum of $j+N_i$ and $i+N_j$. This gives
    \begin{align}
        g_{n'}(m_i^2,m_j^2)=\prod_{n=n'}^\infty\frac{m_i^2-f_n(m_j^2)}{m_j^2-f_n(m_i^2)}=\prod_{n=n'}^\infty\frac{m_i^2-m_{n-j+1}^2}{m_j^2-m_{n-i+1}^2}={\rm finite,\,non}{\rm -}{\rm zero}.
    \end{align} 
    In our proof, the precise value of $n'$ does not matter. We will only use the fact that the tail of the infinite product gives a finite, non-zero contribution. We will drop the subscript of $g$ and the lower limit of the product with the understanding that such that a suitable $n'$ always exists. 

    For simplicity we will only prove that the spacing between the $2nd$ and the $3rd$ pole, $m_3^2-m_2^2$  is the same as the spacing between $1st$ and $2nd$ pole, $m_2^2-m_1^2$. The proof can be generalized to higher poles straightforwardly. We will use the finiteness of $g(m_1^2, m_2^2)$ and  $g(m_1^2, m_3^2)$. Writing explicitly,
    \begin{align}
        g(m_1^2,m_2^2)=\prod_{n}^\infty\frac{m_1^2-m_{n-1}^2}{m_2^2-m_{n}^2},\qquad \tilde g(m_1^2,m_2^2)=\prod_{n}^\infty\frac{m_1^2-m_{n-2}^2}{m_2^2-m_{n-1}^2}.
    \end{align}
    In defining the finite quantity $\tilde g$,  we have simple shifted the dummy index $n$ to $n-1$. Now we consider the finite, non-zero ratio,
    \begin{align}\notag
        \frac{g(m_1^2,m_3^2)}{g(m_1^2,m_2^2)\tilde g(m_1^2,m_2^2)}&=\prod_{n}^\infty\frac{m_1^2-m_{n-2}^2}{m_3^2-m_{n}^2}\frac{m_2^2-m_{n}^2}{m_1^2-m_{n-1}^2}\frac{m_2^2-m_{n-1}^2}{m_1^2-m_{n-2}^2}\notag\\
        &=\prod_{n}^\infty\frac{1-\frac{m_2^2}{m_{n-1}^2}}{1-\frac{m_1^2}{m_{n-1}^2}}\times \prod_{n}^\infty \frac{1-\frac{m_2^2}{m_{n}^2}}{1-\frac{m_3^2}{m_{n}^2}}.
    \end{align}
    Shifting the dummy variable $n-1$ to $n$ in the first factor, we get the finite, non-zero quantity,
    \begin{align}
        \hat g &\equiv \prod_{n}^\infty\frac{\Big(1-\frac{m_2^2}{m_{n}^2}\Big)^2}{\Big(1-\frac{m_1^2}{m_{n}^2}\Big)\Big( 1-\frac{m_3^2}{m_{n}^2}\Big)}.\notag\\
        \log(\hat g) &=\sum_{n}^{\infty} 2\log(1-\frac{m_2^2}{m_n^2})-\log(1-\frac{m_3^2}{m_n^2})-\log(1-\frac{m_1^2}{m_n^2}).
    \end{align} 
    Taylor expanding the right hand side,
    \begin{align}
        \log(\hat g)= (m_1^2+m_3^2-2m_2^2)\Big(\sum_{n}^\infty\frac{1}{m_n^2}\Big)-\frac12(m_1^4+m_3^4-2m_2^4)\Big(\sum_{n}^\infty\frac{1}{m_n^4}\Big)+\ldots .
    \end{align}
    The sum is convergent only if term involving  $\sum_n 1/m_n^2$  is convergent as that is the most divergent sum. Now, if we show that $\sum_n 1/m_n^2$ is divergent, finiteness of $\log(\hat g)$ would imply $m_1^2+m_3^2-2m_2^2=0$ i.e. $m_3^2-m_2^2=m_2^2-m_1^2$. 

    We will now proceed to show that the sum $\sum_n 1/m_n^2$ can not be convergent. If the spectrum of poles accumulate then $\sum_n 1/m_n^2$ does not converge leading to a contradiction. So we will assume that the spectrum of poles does not accumulate. In particular we will take $\lim_{N\to \infty} m_N^2=\infty$. Let us assume that the sum $\sum_n 1/m_n^2$ is convergent and show a contradiction. The crossing symmetry implies $N(s,t)=N(t,s)$ i.e.
    \begin{align}
        \prod_n(s-f_n(t))B_n(s,t)= \prod_n(t-f_n(s))B_n(t,s).
    \end{align}
    Computing the left hand side at $t=m_1^2$, we get the finite quantity,
    \begin{align}
        \prod_n (1-\frac{s}{m_n^2})(-m_n^2 B_n(s,m_1^2)).
    \end{align}
    The convergence of sum $\sum_n 1/m_n^2$, implies convergence of the product $\prod_n (1-\frac{s}{m_n^2})$. This implies $\prod_n(-m_n^2)B_n(s,m_1^2)$ is finite. Similarly, evaluating the right hand side at $s=m_2^2$, we get another finite, non-zero quantity, $\prod_n(-m_{n-1}^2)B_n(m_2^2,t)$. Taking their ratio we get another finite quantity,
    \begin{align}
        \alpha(s,t)\equiv \prod_{n} \frac{m_n^2 B_n(s,m_1^2)}{m_{n-1}^2 B_n(m_2^2,t)}
    \end{align}
    Using enhanced crossing symmetry, we have the non-zero finite quantity
    \begin{align}
        \beta\equiv \prod_n \frac{B_n(m_2^2,m_1^2)}{B_n(m_1^2,m_2^2)},
    \end{align}
    Taking the ratio $\alpha(m_2^2,m_1^2)/\beta$ we get the finite, non-zero quantity,
    \begin{align}
        \prod_n \frac{m_n^2}{m_{n-1}^2} = \lim_{N\to \infty}\prod_{n}^{N}\frac{m_n^2}{m_{n-1}^2}=\lim_{N\to \infty} m_N^2/m_n^2.
    \end{align}
    This is a contradiction because $m_N^2\to \infty $ as $N\to \infty$. 
\end{proof}
\begin{shaded*}
\maintheoremtwo*
\end{shaded*}
\begin{proof}
The canonical form of the partially \ie\ S-matrix is
\begin{align}
    \CM(s,t)&=\frac{\prod_{n=1}^\infty A_n(t)(s-f_n(t))B_n(s,t)}{\prod_{n=1} (s-m_n^2) \CC_n(s)\prod_{n=1}^\infty (t-m_n^2)\CC_n(t)}\notag\\
    &=\prod_{n=1}^\infty \frac{A_n(t)(s-f_n(t))}{(s-m_n^2)(t-m_n^2)}K_n(s,t), \qquad K_n(s,t)=\frac{B_n(s,t)}{C_n(s)C_n(t)}.
\end{align}
The enhanced crossing symmetry implies $\prod_{n=1}^\infty K_n(s,t)/K_n(t,s)=1$.
Computing the residue at $t=m_i^2$, we get
\begin{align}
    {\rm Res}_{t=m_i^2} \CM(s,t)=\kappa \prod_{n=1}^\infty\frac{(s-m_{n-i+1}^2)}{(s-m_n^2)}K_n(s,m_i^2),\qquad \kappa=\prod_{n=1}^\infty \frac{A_n(m_i^2)}{C_n(m_i^2)}\prod_{n\neq i}^\infty\frac{1}{(m_i^2-m_n^2)}. \notag
\end{align}
Here $m_j^2$ for negative value of $j$ stands for the position of excess zeros. The ratio
\begin{align}
    \prod_{n=1}^\infty\frac{(s-m_{n-i+1}^2)}{(s-m_n^2)} = \frac{\prod_{n=0}^{i-1}(s-m_{n-i}^2)}{(s-m_\infty^2)^i}.
\end{align}
The residue is a polynomial with zeros at $m_{n-i+1}^2$ for $n=0,\ldots, i-1 $. This implies
\begin{align}
    \prod_{n=1}^\infty K_n(s,m_i^2)=y(m_i^2) (s-m_\infty^2)^i.
\end{align} 
This, along with the enhanced crossing symmetry of $K_n(s,t)$ yields,
\begin{align}\label{ec-cross}
    \prod_{n=1}^\infty \frac{K_n(m_j^2,m_i^2)}{K_n(m_i^2,m_j^2)} =\frac{y(m_i^2)(m_j^2-m_\infty^2)^i}{y(m_j^2)(m_i^2-m_\infty^2)^j}=1.
\end{align}
Setting $j=1$,
\begin{align}
    \frac{y(m_i^2)(m_1^2-m_\infty^2)^i}{y(m_1^2)(m_i^2-m_\infty^2)}=1.
\end{align}
Substituting this value of $y(m_i^2)$ in the enhanced crossing equation \eqref{ec-cross}, we get
\begin{align}
    \frac{1}{(m_i^2-m_\infty^2)^{j-1}(m_1^2-m_\infty^2)^i}&=\frac{1}{(m_j^2-m_\infty^2)^{i-1}(m_1^2-m_\infty^2)^j}\notag\\
    \Rightarrow \quad\frac{(m_1^2-m_\infty^2)^{j-1}}{(m_i^2-m_\infty^2)^{j-1}}&=\frac{(m_1^2-m_\infty^2)^{i-1}}{(m_j^2-m_\infty^2)^{i-1}}.
\end{align}
Taking $1/(i-1)(j-1)$ power on both sides, we see that the ratio
\begin{align}
    \frac{(m_1^2-m_\infty^2)^{1/(i-1)}}{(m_i^2-m_\infty^2)^{1/(i-1)}}
\end{align}
is independent of $i$. This gives us the desired result. 
\end{proof}

\section{Some instructive examples}\label{examples}
Let us call the property of the amplitude where the degree of the residue polynomial increases by $1$ as $t$ moves from one pole to the next one, the degree-increase property.  
In this section, we will take a look at some examples of seemingly consistent classical S-matrices but with no degree-increase property.  
Due to lemma \ref{degree-increase}, this must mean that this S-matrix must not be \ie, not even partially \ie. 
In the analysis of such S-matrices, we will discover an interesting non-isolated singularity of the zero-functions akin to the so-called natural boundary. 
We will also take a look at the example where the S-matrix does have degree-increase property but the poles are not equally spaced.  This must imply that one of the conditions of theorem \ref{main-theorem} must be invalidated. We will see how that happens. 

\subsection{Exchange of infinitely many scalars}\label{sec-inf-scalars}
Perhaps the simplest possible amplitude with infinitely many spins is the one corresponding to exchange of infinitely many scalar particles. It takes the form,
\begin{align}\label{inf-scalars}
    \CM(s,t)=\sum_{i=1}^\infty c_i\Big(\frac{1}{s-m_i^2}+\frac{1}{t-m_i^2}\Big).
\end{align}
Assume that $c_i$ and $m_i^2$ are such that the infinite sum is convergent. We also assume that $\lim_{N\to \infty} m_N^2=\infty$, in other words, the poles of the S-matrix do not accumulate. 
We will give a concrete example of such an S-matrix shortly. 
Clearly, \eqref{inf-scalars} obeys all the defining properties of the S-matrix including positivity and crossing. However, it does not satisfy the degree-increase property. In fact, the degree of the residue polynomial stays the same at all poles. Due to lemma \ref{degree-increase}, this must mean that this S-matrix must not be partially  \ie.
Now we will see how and why that is the case. 

It is instructive to first consider the case of finitely many scalar exchanges. This takes the same form as  \eqref{inf-scalars}, except that the sum is truncated at some $i=N$. As the residue at $t=m_1^2$ is a constant, it must be that at $t=m_1^2$, all the $s$-poles are cancelled by $s$-zeros. As $t$ moves from $m_1^2$ to $m_2^2$, all the zeros must move left to cancel the all the poles once again. This is because the residue at $t=m_2^2$ is also a constant. The zero function $f_i$, $i>1$ must cancel the pole at $s=m_{i-1}^2$. This is because, there is only a single excess zero $f_1$, hence it can not recombine with a partner to become complex. In turn, no pair of complex zeros can come to the real axis and recombine with $f_i$, $i>1$. At $t=m_2^2$, all the poles are canceled in this fashion except for the last pole $s=m_N^2$. For this poles to cancel, the zero function $f_1$ must shoot off to $-\infty$ and come back from the $+\infty$ direction to cancel the pole $s=m_N^2$. This is because, it has to remain real throughout the path that goes from $s=m_1^2$ to $s=m_N^2$ without encountering other poles. The same process repeats as $t$ moves from $m_2^2$ to $m_3^2$ and so on. It is indeed true that the zeros of 
\begin{align}
    \alpha+\sum_{i=1}^\infty c_i\Big(\frac{1}{s-m_i^2}\Big)=0, \qquad \alpha\in {\mathbb R}
\end{align}
are real for any value of $c_i>0$ and $a_i\in {\mathbb R}$. This is because as $t$ varies over $(m_1^2,m_2^2)$, the $t$-dependent part of the S-matrix can take any real value $\alpha$ as there are poles with positive residue both at $t=m_1^2$ and $t=m_2^2$. The same reasoning holds for the $s$-dependent part of the S-matrix. It can take any real value $-\alpha$ as $s$ moves from one pole with positive residue to the next one with positive residue. 

We can think of the S-matrix \eqref{inf-scalars} as a limit $N\to \infty$ of the finite case. Then, as $t$ moves from $m_1^2$ to $m_2^2$, the zero $f_1$ moves to $-\infty$ to cancel the ``last zero'' from the other side. This clearly makes $\CM(s,t)$ in \eqref{inf-scalars} non-\ie. In fact, something more interesting happens. 
In the $N\to \infty$, the last pole $m_N^2$ goes to $+\infty$. As $t$ moves from $m_1^2$ to $m_2^2$, the zero $f_1$ must move from $m_N^2$ to $m_{N-1}^2$. Both these poles are at infinity in the $N\to \infty$ limit. As a result, the zero function is divergent for any real value of $t\geq m_1^2$. 
This produces an exotic type of non-isolated singularity of the zero function. Such non-isolated singularities go under the general name of natural boundary. Natural boundaries have appeared in the study of S-matrices before \cite{Freund1961, wongnatural, 10.1063/1.1704704, PhysRev.138.B187, Mizera:2022dko}. A simple example of a function with natural boundary is the so-called ``lacunary series'' \cite{ bams/1183525927}.
As this boundary contains the poles $m_i^2$, the S-matrix is not partially \ie. 

In order to construct, a concrete example of such an S-matrix, we need only find a family of $c_i$ and $m_i^2$ so that the sum in \eqref{inf-scalars} is convergent. Let us pick $c_i=1$ and $m_i^2=i^2$. Then we get
\begin{align}
    \CM(s,t)=\sum_{i=1}^\infty \Big(\frac{1}{s-i^2}+\frac{1}{t-i^2}\Big) =\frac{\pi \sqrt{s} \cot(\pi \sqrt{s})-1}{2s} +\frac{\pi \sqrt{t} \cot(\pi \sqrt{t})-1}{2t}. 
\end{align}
We can also produce an example where the poles of the S-matrix accumulate. In this case, the role of $\infty$ where the zeros get ``stuck'' as $t$ takes successive pole values is played by $m_\infty^2$. As a result, the zero functions have a natural boundary like singularity where the functions $f_i(t)$ takes a constant value for real $t$ greater than a certain value. To the best of our knowledge, such singularities have not appeared in maths literature.
To produce a concrete example, we take $c_i=1/i^2$ and $m_i=1-1/i$. Then we get,
\begin{align}
    \CM(s,t)=\sum_{i=1}^\infty \frac{1}{i^2}\Big(\frac{1}{s-(1-\frac{1}{i})}+\frac{1}{t-(1-\frac{1}{i})}\Big)= 2\gamma+\psi\Big(\frac{s}{s-1}\Big)+\psi\Big(\frac{t}{t-1}\Big),
\end{align}
where $\psi(z)\equiv d_z\log\Gamma(z)$, known as the digamma function. 
This extended singularity for the zero functions is somewhat of a generic phenomenon for non-\ie\ S-matrices as we illustrate more below. 

\subsection{Sum of Veneziano-type amplitudes}
Consider a generalization of the Veneziano amplitude
\begin{align}
    \CM_{ij}(s,t)=\frac{\Gamma(i-s)\Gamma(i-t)}{\Gamma(j-s-t)}.
\end{align}
In this notation, the Veneziano amplitude is $\CM_{-1,-2}(s,t)$. Consider the sum,
\begin{align}
    \CM(s,t)=\CM_{0,0}(s,t)+\CM_{\frac12,1}(s,t).
\end{align}
This has a tower of poles spaced by $\frac12$.  However, it does not satisfy the degree-increase property. The degree of the residue polynomial  rather behaves as $\lfloor\frac{i}{2}\rfloor$. Due to lemma \ref{degree-increase}, this S-matrix can not \ie. In fact it can not even be partially \ie. Now we will see how that happens. 

The situation is quite similar to the one discussed in section \ref{sec-inf-scalars}. It is instructive to take $t=\frac12-\epsilon$ with $\epsilon$ small and positive. At $t=\frac12$, $\CM_{0,0}$ is finite but $\CM_{\frac12, 1}$ has a pole. Approximating the amplitude for large negative $s$, 
\begin{align}\label{near-half}
    \CM(-|s|,\frac12-\epsilon)= \Gamma(-\frac12) (|s|^\frac12)+\frac1\epsilon,
\end{align}  
As $\Gamma(-\frac12)<0$, we see that one of the zeros of $\CM$, $f_1(t)$, shoots off to $-\infty$ as $-1/\epsilon^2$ as $t$ approaches $1/2$ from the left. In fact, this behavior holds as $t$ approaches any half-integer from the left. 

As $t$ passes $\frac12$ i.e. for $\epsilon < 0$, both the terms in equation \eqref{near-half} are negative so there is no zero on the negative real axis.  
This remains to be the case until $t$ reaches $1$. In other words, the zero $f_1$ remains ``stuck'' at $\infty$ for $t\in [\frac12,1]$. 
As $t$ crosses $1$, the zero $f_1$ again emerges from $-\infty$ direction. It disappears to infinity again as $t\to \frac32$, emerging back at $t=2$ and so on. In this way the smallest zero $f_1$ keeps playing hide and seek.
The set $[\frac12,1]\cup [\frac32,2]\cup\ldots$ forms the natural boundary of $f_1(t)$. As the natural boundary of $f_1$ contains the poles, the S-matrix is not partially \ie. 

The rest of the zero simply gather in the complex plane after they cross the smallest pole successively and remain finite for all values of $t>0$. This picture explains the degree of the residue polynomial at $s=i$ to be $\lfloor\frac{i}{2}\rfloor$.

\subsection{Bespoke Amplitudes}
An elegant approach to constructing amplitudes with a tunable spectrum is the so-called bespoke amplitudes \cite{Cheung:2023Bes}. In this approach, given an S-matrix $\CM(s,t)$, a new S-matrix is constructed as
\begin{align}
    \CM_{\rm bespoke}(s,t)= \sum_{\alpha(s), \alpha(t)} \CM(\alpha(s),\alpha(t)),
\end{align}
where $\alpha(x)$ are all the roots of the equation
\begin{align}
    x=\frac{P(\alpha)}{Q(\alpha)}, \qquad \qquad P(\alpha),Q(\alpha): {\rm polynomials}.
\end{align}
Due to the sum over all the roots, the resulting S-matrix does not have a branch cut. Its analytic structure is as excepted of an S-matrix. However, unitarity is not necessarily preserved by this construction. In general, the degree-increase property is also not preserved by the construction. 
\subsection*{Example 1}
We first will consider a simple example of a bespoke amplitude constructed from the Veneziano amplitude that does not obey the degree-increase property and show that it is consistent with  lemma \ref{degree-increase}.  Consider the bespoke amplitude for $P(\alpha)=\alpha^2, Q(\alpha)=1$.   
\begin{align}
    \CM(s, t) = \frac{\Gamma(-\sqrt{s})\Gamma(-\sqrt{t})}{\Gamma(-\sqrt{s} - \sqrt{t})} + \frac{\Gamma(\sqrt{s})\Gamma(-\sqrt{t})}{\Gamma(\sqrt{s} - \sqrt{t})} + \frac{\Gamma(-\sqrt{s})\Gamma(\sqrt{t})}{\Gamma(-\sqrt{s} + \sqrt{t})} + \frac{\Gamma(\sqrt{s})\Gamma(\sqrt{t})}{\Gamma(\sqrt{s} + \sqrt{t})}
\end{align}
The poles are at $m^{2}_{i} = i^{2}$ and the  order of the residue polynomial is  $ \left\lfloor \frac{i}{2} \right\rfloor $. In particular, as $t$ changes from $m_2^2$ to $m_3^2$, the degree of the residue polynomial does not change. This means that one the zero must shoot off to $\infty$ in this interval. 

This can be seen by examining the S-matrix near $m_2^2$ and $m_3^2$. In this limit, the last two term fall off as $s^{-\sqrt{t}}$ at large $s$. We see numerically that the zero of 
\begin{align}
    \frac{\Gamma(-\sqrt{s})}{\Gamma(-\sqrt{s} - \sqrt{t})} + \frac{\Gamma(\sqrt{s})}{\Gamma(\sqrt{s} - \sqrt{t})}
\end{align}
indeed goes off to infinity as $t$ goes from $m_2^2$ to $m_3^2$. 

\subsection*{Example 2}
Now we will consider another example of the bespoke amplitude that does obey degree-increase theorem but does not have equally spaced spectrum of poles. We take $P(\alpha)=\alpha^2, Q(\alpha)=2(1+\alpha)$. Solving for $\alpha$ we get
\begin{align}
    \alpha_\pm (s)=s\pm \sqrt{s(s+2)}. 
\end{align}
Note that in the large $s$ limit, $\alpha_+(s)\to 2s, \alpha_-(s)\to -1$. For $t=-\epsilon$ and large $s$, the amplitude simplifies to
\begin{align}
    \CM(s,-\epsilon)&=\frac{\Gamma(-2s)\Gamma(-i\sqrt{2\epsilon})}{\Gamma(-2s-i\sqrt{2\epsilon})}+\frac{\Gamma(1)\Gamma(-i\sqrt{2\epsilon})}{\Gamma(1-i\sqrt{2\epsilon})}+ \frac{\Gamma(-2s)\Gamma(i\sqrt{2\epsilon})}{\Gamma(-2s+i\sqrt{2\epsilon})}+\frac{\Gamma(1)\Gamma(i\sqrt{2\epsilon})}{\Gamma(1+i\sqrt{2\epsilon})}
\end{align}
The third and the fourth term goes to zero. Using $\Gamma(i\sqrt{2\epsilon})=\frac{-i}{\sqrt{2\epsilon}}-\gamma$, where $\gamma$ is the Euler-Mascheroni constant. In the large $s$ limit, 
\begin{align}
    \CM(s,t)\to \frac{-i}{\sqrt{2\epsilon}}((2s)^{-i\sqrt{2\epsilon}}-(2s)^{i\sqrt{2\epsilon}})-\gamma((2s)^{-i\sqrt{2\epsilon}}+(2s)^{i\sqrt{2\epsilon}}).
\end{align}
Approximate location of zeros is given by the equation
\begin{align}
    \tan(\sqrt{2\epsilon} \log(2s))=-2\sqrt{2\epsilon}\gamma\qquad \Rightarrow \qquad \log(2s)=-\gamma+\frac{2\pi n}{\sqrt{2\epsilon}}.
\end{align}
These zeros shoot off to infinity as $\epsilon \to 0$. This shows that the S-matrix is not \ie\ as expected from theorem \ref{main-theorem}. 

\subsection{S-matrix with increasing degree of the residue polynomial}
We will now construct another S-matrix which does have the degree-increase property but  does not have an equally spaced spectrum of poles. Due to theorem \ref{main-theorem}, it must be that such an S-matrix is either not \ie, not \ec, or both. We will see how that happens. 

Consider
\begin{align}
    \CM(s,t)= \frac{1}{t} + \frac{1}{s} + \sum_{i=1}^{n} \left( \frac{P_{i}(t)}{s - i^2} + \frac{P_{i}(s)}{t - i^2} \right)
\end{align}
where $P_{i}(t) = \prod_{j=1}^{i} \left( 1 + \frac{t}{j^2} \right)$. Clearly, this S-matrix does obey $d_{i+1}=d_i+1$ i.e. the degree of the residue polynomial increase by $1$. We will analyze the amplitude at $t=-\epsilon$. At leading order we have,
\begin{align}
    \CM(s, -\epsilon) = \frac{1}{-\epsilon} + \frac{1}{s} + \sum_{i=1}^{n} \left( \frac{1}{s - i^2} + \frac{P_{i}(s)}{-i^2} \right).
\end{align}
At large $s$, this approximates to
\begin{align}
    \CM(s, -\epsilon) = \frac{1}{-\epsilon} - \left( 1 + s \right) \, _5F_4\left(\begin{array}{c}
        1, 1, 1, 2 - i \sqrt{s}, 2 + i \sqrt{s} \\
        2, 2, 2, 2
        \end{array}; 1\right),
\end{align}
The second term on the right hand side, is monotonically increasing as $s\to -\infty$. This implies that as $\epsilon\to 0$, one of the zeros must shoot off to $\infty$ along the negative real axis. This makes the amplitude non-\ie. 

\section*{Acknowledgement}

We are grateful to Alok Laddha, Shiraz Minwalla, Sabyasachi Mukherjee,  Ashoke Sen and Sasha Zhiboedov for interesting discussion.  We would also like to thank Shiraz Minwalla and Yu-tin Huang for useful comments on the manuscript. SJ would like to thank Sonal Dhingra and Shivansh Tomar for discussion.  We would like to acknowledge the support of the Department of Atomic Energy, Government of India, under Project Identification No. RTI 4002. This work is partially  supported by the Infosys Endowment for the study of the Quantum Structure of Spacetime. The authors would also like to acknowledge their debt to the people of India for their steady support to the study of the basic sciences.

\bibliography{LargeDCFT}

\end{document}